\documentclass[10pt,journal,epsfig]{IEEEtran}

\usepackage[dvips]{graphicx}
\usepackage{graphicx}
\usepackage{amssymb}
\usepackage{cite}
\usepackage{subfigure}

\usepackage{amsmath}
\usepackage{multirow}
\usepackage{xcolor}
\usepackage[font=small]{caption}

\begin{document}

\title{Low-Rank Tensor
Decomposition-Aided Channel Estimation for Millimeter Wave
MIMO-OFDM Systems}

\author{Zhou Zhou, Jun Fang, Linxiao Yang, Hongbin Li, Zhi Chen, and Rick S. Blum,~\IEEEmembership{Fellow,~IEEE}
\thanks{Zhou Zhou, Jun Fang, Linxiao Yang, Zhi Chen are with the National Key Laboratory
of Science and Technology on Communications, University of
Electronic Science and Technology of China, Chengdu 611731, China,
Email: JunFang@uestc.edu.cn}
\thanks{Hongbin Li is
with the Department of Electrical and Computer Engineering,
Stevens Institute of Technology, Hoboken, NJ 07030, USA, E-mail:
Hongbin.Li@stevens.edu}
\thanks{Rick S. Blum is with the Department of Electrical and Computer Engineering, Lehigh University,
Bethlehem, PA 18015, USA, E-mail: rblum@lehigh.edu}
\thanks{This work was supported in part by the National Science
Foundation of China under Grant 61522104, the National Science
Foundation under Grants ECCS-1408182 and ECCS-1609393, and the Air
Force Office of Scientific Research under grant
FA9550-16-1-0243.}}

\maketitle






\begin{abstract}
We consider the problem of downlink channel estimation for
millimeter wave (mmWave) MIMO-OFDM systems, where both the base
station (BS) and the mobile station (MS) employ large antenna
arrays for directional precoding/beamforming. Hybrid analog and
digital beamforming structures are employed in order to offer a
compromise between hardware complexity and system performance.
Different from most existing studies that are concerned with
narrowband channels, we consider estimation of wideband mmWave
channels with frequency selectivity, which is more appropriate for
mmWave MIMO-OFDM systems. By exploiting the sparse scattering
nature of mmWave channels, we propose a CANDECOMP/PARAFAC (CP)
decomposition-based method for channel parameter estimation
(including angles of arrival/departure, time delays, and fading
coefficients). In our proposed method, the received signal at the
BS is expressed as a third-order tensor. We show that the tensor
has the form of a low-rank CP decomposition, and the channel
parameters can be estimated from the associated factor matrices.
Our analysis reveals that the uniqueness of the CP decomposition
can be guaranteed even when the size of the tensor is small. Hence
the proposed method has the potential to achieve substantial
training overhead reduction. We also develop Cram\'{e}r-Rao bound
(CRB) results for channel parameters, and compare our proposed
method with a compressed sensing-based method. Simulation results
show that the proposed method attains mean square errors that are
very close to their associated CRBs, and presents a clear
advantage over the compressed sensing-based method in terms of
both estimation accuracy and computational complexity.
\end{abstract}



\section{Introduction}
Millimeter-wave (mmWave) communication is a promising technology
for future cellular networks
\cite{RanganRappaport14,GhoshThomas14}. The large bandwidth
available in mmWave bands can offer gigabit-per-second
communication data rates. However, high signal attenuation at such
high frequency presents a major challenge for system design
\cite{SwindlehurstAyanoglu14}. To compensate for the significant
path loss, large antenna arrays should be used at both the base
station (BS) and the mobile station (MS) to provide sufficient
beamforming gain for mmWave communications \cite{AlkhateebMo14}.
This requires accurate channel estimation which is essential for
the proper operation of directional precoding/beamforming in
mmWave systems.

Channel estimation in mmWave systems is challenging due to hybrid
precoding structures and the large number of antennas. A primary
challenge is that hybrid precoding structures
\cite{AyachRajagopal14,AlkhateebLeus15b,GaoDai15,KulkarniGhosh16}
employed in mmWave systems prevent the digital baseband from
directly accessing the entire channel dimension. This is also
referred to as the channel subspace sampling limitation
\cite{HurKim13,KimLove15}, which makes it difficult to acquire
useful channel state information (CSI) during a practical channel
coherence time. To address this issue, fast beam scanning and
searching techniques have been extensively studied, e.g.
\cite{HurKim13,Wang09,TsangPoon11}. The objective of beam scanning
is to search for the best beamformer-combiner pair by letting the
transmitter and receiver scan the adaptive sounding beams or coded
beams chosen from pre-determined sounding beam codebooks.
Nevertheless, as the number of antennas increases, the size of the
codebook should be enlarged accordingly, which in turn results in
an increase in the sounding/training overhead.

Unlike beam scanning techniques whose objective is to find the
best beam pair, another approach is to directly estimate the
mmWave channel or its associated parameters, e.g.
\cite{AlkhateebyLeus15,AlkhateebAyach14,SchniterSayeed14,KimLove15,ZhouFang16}.
In particular, by exploiting the sparse scattering nature of the
mmWave channels, mmWave channel estimation can be formulated as a
sparse signal recovery problem, and it has been shown
\cite{AlkhateebyLeus15,AlkhateebAyach14} that substantial
reduction in training overhead can be achieved via compressed
sensing methods. In \cite{AlkhateebAyach14}, an adaptive
compressed sensing method was developed for mmWave channel
estimation based on a hierarchical multi-resolution beamforming
codebook. Compared to the standard compressed sensing method, the
adaptive method is more efficient as the training precoding is
adaptively adjusted according to the outputs of earlier stages.
Nevertheless, this improved efficiency comes at the expense of
requiring feedback from the MS to the BS. Other compressed
sensing-based mmWave channel estimation methods include
\cite{RamasamyVenkateswaran12a,RamasamyVenkateswaran12b,MarziRamasamy16,XieGao16}.
Most of the above existing methods are concerned with estimation
of narrowband channels. MmWave systems, however, are very likely
to operate on wideband channels with frequency selectivity
\cite{AlkhateebHeath16}. In \cite{GaoHu16}, the authors considered
the problem of multi-user uplink channel estimation in mmWave
MIMO-OFDM systems and proposed a distributed compressed
sensing-based scheme by exploiting the angular domain structured
sparsity of mmWave wideband frequency-selective fading channels.
Precoding design, with limited feedback for frequency selective
wideband mmWave channels, was studied in \cite{AlkhateebHeath16}.




In this paper, we study the problem of downlink channel estimation
for mmWave MIMO-OFDM systems, where wideband frequency-selective
fading channels are considered. We propose a CANDECOMP/PARAFAC
(CP) decomposition-based method for downlink channel estimation.
The proposed method is based on the following three key
observations. First, by adopting a simple setup at the
transmitter, the received signal at the BS can be organized into a
third-order tensor which admits a CP decomposition. Second, due to
the sparse scattering nature of mmWave channels, the tensor has an
intrinsic low CP rank that guarantees the uniqueness of the CP
decomposition. Third, the channel parameters, including angles of
arrival/departure, time delays, and fading coefficients, can be
easily extracted based on the decomposed factor matrices. We
conduct a rigorous analysis on the uniqueness of the CP
decomposition. Analyses show that the uniqueness of the CP
decomposition can be guaranteed even when the size of the tensor
is small. This result implies that our proposed method can achieve
a substantial training overhead reduction. The Cram\'{e}r-Rao
bound (CRB) results for channel parameters are also developed,
which provides a benchmark for the performance of our proposed
method, and also describes the best asymptotically achievable
performance. Our experiments show that the mean square errors
attained by the proposed method are close to their corresponding
CRBs.

Our proposed CP decomposition-based method enjoys the following
advantages as compared with the compressed sensing-based method.
Firstly, unlike compressed sensing techniques which require to
discretize the continuous parameter space into a finite set of
grid points, our proposed method is essentially a gridless
approach and therefore is free of the grid discretization errors.
Secondly, the proposed method captures the intrinsic
multi-dimensional structure of the multiway data, which helps
achieve a performance improvement. Thirdly, the use of tensors for
data representation and processing leads to a very low
computational complexity, whereas most compressed sensing methods
are usually plagued by high computational complexity. Our
simulation results show that our proposed method has a
computational complexity as low as the simplest compressed sensing
method, i.e. the orthogonal matching pursuit (OMP) method
\cite{PatiRezaiifar93}, while achieving a much higher estimation
accuracy than the OMP. Lastly, the conditions for the uniqueness
of the CP decomposition are easy to analyze, and can be employed
to determine the exact amount of training overhead required for
unique decomposition. In contrast, it is usually difficult to
analyze and check the exact recovery condition for generic
dictionaries for compressed sensing techniques.



The rest of the paper is organized as follows. In Section
\ref{sec:preliminaries}, we provide notations and basics on the CP
decomposition. The system model and the channel estimation problem
are discussed in Section \ref{sec:system-model}. In Section
\ref{sec:proposed-method}, we propose a CP decomposition-based
method for mmWave channel estimation. The uniqueness of the CP
decomposition is also analyzed. Section \ref{sec:CRB} develops CRB
results for the estimation of channel parameters. A compressed
sensing-based channel estimation method is discussed in Section
\ref{sec:cs-method}. Computational complexity of the proposed
method and the compressed sensing-based method is analyzed in
Section \ref{sec:complexity-analysis}. Simulation results are
provided in Section \ref{sec:experiments}, followed by concluding
remarks in Section \ref{sec:conclusion}.

\begin{figure}[!t]
\centering
\includegraphics[width=9cm]{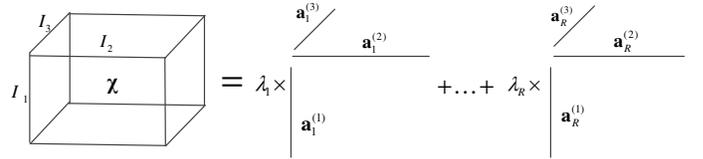}
\caption{Schematic of CP decomposition.} \label{fig1}
\end{figure}

\begin{figure*}[!t]
\centering
\includegraphics[width=16.4cm]{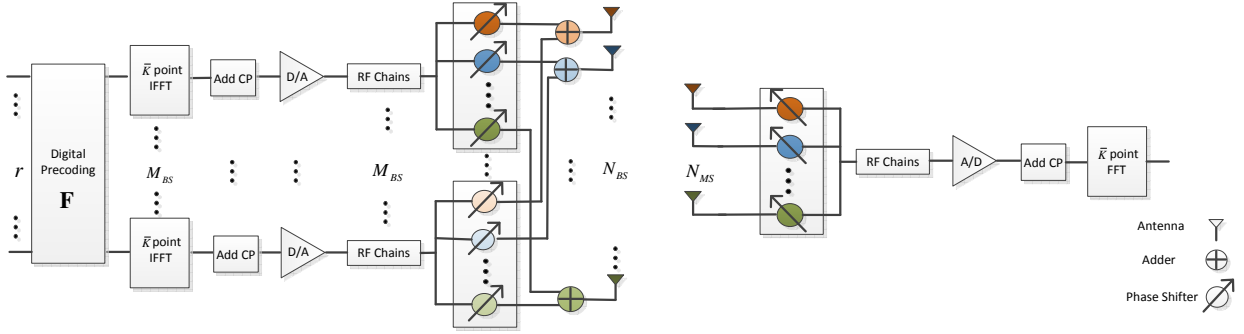}
\caption{A block diagram of the MIMO-OFDM transceiver that employs
hybrid analog/digital precoding.} \label{fig8}
\end{figure*}

\section{Preliminaries} \label{sec:preliminaries}
To make the paper self-contained, we provide a brief review on
tensors and the CP decomposition. More details regarding the
notations and basics on tensors can be found in
\cite{KoldaBader09,Kolda06,CichockiMandic15}. Simply speaking, a
tensor is a generalization of a matrix to higher-order dimensions,
also known as ways or modes. Vectors and matrices can be viewed as
special cases of tensors with one and two modes, respectively.
Throughout this paper, we use symbols $\otimes$ , $\circ$ , and
$\odot$ to denote the Kronecker, outer, and Khatri-Rao product,
respectively.

Let $\boldsymbol{\mathcal{X}}\in\mathbb{C}^{I_1\times
I_2\times\cdots\times I_N}$ denote an $N$th-order tensor with its
$(i_1,\ldots,i_N)$th entry denoted by $\mathcal{X}_{i_1\cdots
i_N}$. Here the order $N$ of a tensor is the number of dimensions.
Fibers are a higher-order analogue of matrix rows and columns. The
mode-$n$ fibers of $\boldsymbol{\mathcal{X}}$ are
$I_n$-dimensional vectors obtained by fixing every index but
$i_n$. Slices are two-dimensional sections of a tensor, defined by
fixing all but two indices. Unfolding or matricization is an
operation that turns a tensor into a matrix. The mode-$n$
unfolding of a tensor $\boldsymbol{\mathcal{X}}$, denoted as
$\boldsymbol{X}_{(n)}$, arranges the mode-$n$ fibers to be the
columns of the resulting matrix. The CP decomposition decomposes a
tensor into a sum of rank-one component tensors (see Fig.
\ref{fig1}), i.e.
\begin{align}
\boldsymbol{\mathcal{X}}=
\sum\limits_{r=1}^{R}\lambda_r\boldsymbol{a}_r^{(1)}\circ\boldsymbol{a}_r^{(2)}\circ\cdots\circ\boldsymbol{a}_r^{(N)}
\end{align}
where $\boldsymbol{a}_r^{(n)}\in\mathbb{C}^{I_n}$, the minimum
achievable $R$ is referred to as the rank of the tensor, and
$\boldsymbol{A}^{(n)}\triangleq
[\boldsymbol{a}_{1}^{(n)}\phantom{0}\ldots\phantom{0}\boldsymbol{a}_{R}^{(n)}]\in\mathbb{C}^{I_n\times
R}$ denotes the factor matrix along the $n$-th mode. Elementwise,
we have
\begin{align}
\mathcal{X}_{i_1 i_2\cdots i_N}=\sum\limits_{r=1}^{R}\lambda_r
a_{i_1 r}^{(1)}a_{i_2 r}^{(2)}\cdots a_{i_N r}^{(N)}
\end{align}
The mode-$n$ unfolding of $\boldsymbol{\mathcal{X}}$ can be
expressed as
\begin{align}
\boldsymbol{X}_{(n)}=\boldsymbol{A}^{(n)}\boldsymbol{\Lambda}\left(\boldsymbol{A}^{(N)}
\odot\cdots\boldsymbol{A}^{(n+1)}\odot\boldsymbol{A}^{(n-1)}\odot\cdots\boldsymbol{A}^{(1)}\right)^T
\end{align}
where
$\boldsymbol{\Lambda}\triangleq\text{diag}(\lambda_1,\ldots,\lambda_R)$.



\section{System Model} \label{sec:system-model}
Consider a mmWave massive MIMO-OFDM system consisting of a base
station (BS) and multiple mobile stations (MSs). To facilitate the
hardware implementation, hybrid analog and digital beamforming
structures are employed by both the BS and the MS. We assume that
the BS is equipped with $N_{\text{BS}}$ antennas and
$M_{\text{BS}}$ RF chains, and the MS is equipped with
$N_{\text{MS}}$ antennas and $M_{\text{MS}}$ RF chains. The number
of RF chains is less than the number of antennas, i.e.
$M_{\text{BS}}<N_{\text{BS}}$ and $M_{\text{MS}}<N_{\text{MS}}$.
In particular, we assume $M_{\text{MS}}=1$, i.e. each MS has only
one RF chain. The total number of OFDM tones (subcarriers) is
assumed to be $\bar{K}$, among which $K$ subcarriers are selected
for training. For simplicity, here we assume subcarriers
$\{1,2,\ldots,K\}$ are assigned for training. Nevertheless, our
formulation and method can be easily extended to other subset
choices. In the downlink scenario, we only need to consider a
single user system because the channel estimation is conducted by
each user individually.





We adopt a downlink training scheme similar to
\cite{AlkhateebAyach14,AlkhateebyLeus15}. For each subcarrier, the
BS employs $T$ different beamforming vectors at $T$ successive
time frames. Each time frame is divided into $M$ sub-frames, and
at each sub-frame, the MS uses an individual combining vector to
detect the transmitted signal. The beamforming vector associated
with the $k$th subcarrier at the $t$th time frame can be expressed
as
\begin{align}
\boldsymbol{x}_k(t)=\boldsymbol{F}_{\text{RF}}(t)\boldsymbol{F}_k(t)\boldsymbol{s}_k(t)
\quad \forall k=1,\ldots,K\label{eqn2}
\end{align}
where $\boldsymbol{s}_k(t)\in\mathbb{C}^{r}$ denotes the pilot
symbol vector,
$\boldsymbol{F}_k(t)\in\mathbb{C}^{M_{\text{BS}}\times r}$ denotes
the digital precoding matrix for the $k$th subcarrier, and
$\boldsymbol{F}_{\text{RF}}(t)\in\mathbb{C}^{N_{\text{BS}}\times
M_{\text{BS}}}$ is a common RF precoder for all subcarriers. The
procedure to generate the beamforming vector (\ref{eqn2}) is
elaborated as follows. The pilot symbol vector
$\boldsymbol{s}_k(t)$ at each subcarrier is first precoded using a
digital precoding matrix $\boldsymbol{F}_k(t)$. The symbol blocks
are transformed to the time-domain using $\bar{K}$-point inverse
discrete Fourier transform (IDFT). A cyclic prefix is then added
to the symbol blocks, finally a common RF precoder
$\boldsymbol{F}_{\text{RF}}(t)$ is applied to all subcarriers.

At each time frame, the MS successively employs $M$ RF combining
vectors $\{\boldsymbol{q}_m\}$ to detect the transmitted signal.
Note that these combining vectors are common to all subcarriers.
At each sub-frame, the received signal is first combined in the RF
domain. Then, the cyclic prefix is removed and symbols are
converted back to the frequency domain by performing a discrete
Fourier transform (DFT). After processing, the received signal
associated with the $k$th subcarrier at the $m$th sub-frame can be
expressed as \cite{AlkhateebHeath16}
\begin{align}
y_{k,m}(t)=\boldsymbol{q}_m^T
\boldsymbol{H}_k\boldsymbol{x}_k(t)+w_{k,m}(t)
\end{align}
where $\boldsymbol{q}_m\in\mathbb{C}^{N_{\text{MS}}}$ denotes the
combining vector used at the $m$th sub-frame,
$\boldsymbol{H}_k\in\mathbb{C}^{N_{\text{MS}}\times
N_{\text{BS}}}$ is the channel matrix associated with the $k$th
subcarrier, and $w_{k,m}(t)$ denotes the additive Gaussian noise.
Collecting the $M$ received signals $\{y_{k,m}(t)\}_{m=1}^M$ at
each time frame, we have
\begin{align}
\boldsymbol{y}_{k}(t)=&\boldsymbol{Q}^T
\boldsymbol{H}_k\boldsymbol{x}_k(t)+\boldsymbol{w}_k(t)
\nonumber\\
=&\boldsymbol{Q}^T \boldsymbol{H}_k
\boldsymbol{F}_{\text{RF}}(t)\boldsymbol{F}_k(t)\boldsymbol{s}_k(t)+\boldsymbol{w}_k(t)
\end{align}
where
\begin{align}
\boldsymbol{y}_{k}(t)\triangleq &
[y_{k,1}(t)\phantom{0}\ldots\phantom{0}y_{k,M}(t)]^T \nonumber\\
\boldsymbol{w}_{k}(t)\triangleq &
[w_{k,1}(t)\phantom{0}\ldots\phantom{0}w_{k,M}(t)]^T \nonumber\\
\boldsymbol{Q}\triangleq &
[\boldsymbol{q}_1\phantom{0}\ldots\phantom{0}\boldsymbol{q}_M]
\end{align}

Measurement campaigns in dense-urban NLOS environments reveal that
mmWave channels typically exhibit limited scattering
characteristics \cite{AkdenizLiu14}. Also, considering the
wideband nature of mmWave channels, we adopt a geometric wideband
mmWave channel model with $L$ scatterers between the MS and the
BS. Each scatterer is characterized by a time delay $\tau_l$,
angles of arrival and departure (AoA/AoD),
$\theta_l,\phi_l\in[0,2\pi]$. With these parameters, the channel
matrix in the delay domain can be written as
\cite{AlkhateebHeath16,GaoHu16}
\begin{align}
\boldsymbol{H}(\tau)=\sum_{l=1}^L \alpha_l
\boldsymbol{a}_{\text{MS}}(\theta_{l})\boldsymbol{a}_{\text{BS}}^T(\phi_{l})\delta(\tau-\tau_l)
\end{align}
where $\alpha_l$ is the complex path gain associated with the
$l$th path, $\boldsymbol{a}_{\text{MS}}(\theta_{l})$ and
$\boldsymbol{a}_{\text{BS}}(\phi_{l})$ are the antenna array
response vectors of the MS and BS, respectively, and
$\delta(\cdot)$ represents the delta function. Throughout this
paper, we assume
\begin{itemize}
\item[A1] Different scatterers have different angles of
arrival, angles of departure as well as time delays, i.e.
$\theta_i\neq\theta_j$, $\phi_i\neq\phi_j$, $\tau_i\neq\tau_j$ for
$i\neq j$.
\end{itemize}
Since scatterers are randomly distributed in space, this
assumption is usually valid in practice.


Given the delay-domain channel model, the frequency-domain channel
matrix $\boldsymbol{H}_k$ associated with the $k$th subcarrier can
be obtained as
\begin{align}
\boldsymbol{H}_k = \sum_{l=1}^L \alpha_l \exp(-j2\pi\tau_l f_s
k/\bar{K})
\boldsymbol{a}_{\text{MS}}(\theta_{l})\boldsymbol{a}_{\text{BS}}^T(\phi_{l})
\label{Hk}
\end{align}
where $f_s$ denotes the sampling rate. Our objective is to
estimate the channel matrices
$\{\boldsymbol{H}_k\}_{k=1}^{\bar{K}}$ from the received signals
$\boldsymbol{y}_k(t),\forall k=1,\ldots,K,\forall t=1,\ldots,T$.
In particular, we wish to provide a reliable channel estimate by
using as few measurements as possible because the number of
measurements is linearly proportional to the number of time frames
and the number of sub-frames, both of which are expected to be
minimized. To facilitate our algorithm development, we assume that
the digital precoding matrices and the pilot symbols remain the
same for different subcarriers, i.e.
$\boldsymbol{F}_k(t)=\boldsymbol{F}(t)$,
$\boldsymbol{s}_k(t)=\boldsymbol{s}(t),\forall k=1,\ldots,K$. As
will be shown later, this simplification enables us to develop an
efficient tensor factorization-based method to extract the channel
state information from very few number of measurements.


\section{Proposed CP Decomposition-Based Method} \label{sec:proposed-method}
Suppose $\boldsymbol{F}_k(t)=\boldsymbol{F}(t)$ and
$\boldsymbol{s}_k(t)=\boldsymbol{s}(t),\forall k=1,\ldots,K$. Let
$\boldsymbol{S}\triangleq
[\boldsymbol{s}(1)\phantom{0}\ldots\phantom{0}\boldsymbol{s}(T)]$.
The received signal at the $k$th subcarrier can be written as
\begin{align}
\boldsymbol{Y}_k=\boldsymbol{Q}^T
\boldsymbol{H}_k\boldsymbol{P}+\boldsymbol{W}_k \quad
k=1,\ldots,K\label{Yk}
\end{align}
where
\begin{align}
\boldsymbol{Y}_k\triangleq &
[\boldsymbol{y}_k(1)\phantom{0}\ldots\phantom{0}\boldsymbol{y}_k(T)]
\nonumber\\
\boldsymbol{W}_k\triangleq &
[\boldsymbol{w}_k(1)\phantom{0}\ldots\phantom{0}\boldsymbol{w}_k(T)]
\nonumber\\
\boldsymbol{P}\triangleq &
[\boldsymbol{p}(1)\phantom{0}\ldots\phantom{0}\boldsymbol{p}(T)]
\label{P-definition}
\end{align}
in which
$\boldsymbol{p}(t)\triangleq\boldsymbol{F}_{\text{RF}}\boldsymbol{F}(t)\boldsymbol{s}(t)$.

Since signals from multiple subcarriers are available at the MS,
the received signal can be expressed by a third-order tensor
$\boldsymbol{\mathcal{Y}}\in\mathbb{C}^{T\times M\times K}$ whose
three modes respectively stand for the time frame, the sub-frame,
and the subcarrier, and its $(t,m,k)$th entry is given by
$y_{k,m}(t)$. Substituting (\ref{Hk}) into (\ref{Yk}), we obtain
\begin{align}
\boldsymbol{Y}_k=&\sum_{l=1}^L\tilde{\alpha}_{l,k}\boldsymbol{Q}^T
\boldsymbol{a}_{\text{MS}}(\theta_{l})\boldsymbol{a}_{\text{BS}}^T(\phi_{l})
\boldsymbol{P} +\boldsymbol{W}_k \nonumber\\
=& \sum_{l=1}^L\tilde{\alpha}_{l,k}
\boldsymbol{\tilde{a}}_{\text{MS}}(\theta_{l})\boldsymbol{\tilde{a}}_{\text{BS}}^T(\phi_{l})
+\boldsymbol{W}_k
\end{align}
where $\tilde{\alpha}_{l,k}\triangleq\alpha_l\exp(-j2\pi\tau_l f_s
k/\bar{K})$,
$\boldsymbol{\tilde{a}}_{\text{MS}}(\theta_{l})\triangleq\boldsymbol{Q}^T
\boldsymbol{a}_{\text{MS}}(\theta_{l})$, and
$\boldsymbol{\tilde{a}}_{\text{BS}}(\phi_{l})\triangleq\boldsymbol{P}^T\boldsymbol{a}_{\text{BS}}(\phi_{l})$.
We see that each slice of the tensor $\boldsymbol{\mathcal{Y}}$,
$\boldsymbol{Y}_k$, is a weighted sum of a common set of rank-one
outer products. The tensor $\boldsymbol{\mathcal{Y}}$ thus admits
the following CANDECOMP/PARAFAC (CP) decomposition which
decomposes a tensor into a sum of rank-one component tensors, i.e.
\begin{align}
\boldsymbol{\mathcal{Y}}=\sum_{l=1}^L\boldsymbol{\tilde{a}}_{\text{MS}}(\theta_{l})\circ
\boldsymbol{\tilde{a}}_{\text{BS}}(\phi_{l})\circ\boldsymbol{c}_l
+ \boldsymbol{\mathcal{W}} \label{Y-tensor}
\end{align}
where
\begin{align}
\boldsymbol{c}_l\triangleq
[\tilde{\alpha}_{l,1}\phantom{0}\cdots\phantom{0}\tilde{\alpha}_{l,K}]^T=\alpha_l
\boldsymbol{g}(\tau_l)
\end{align}
in which
\begin{align}
\boldsymbol{g}(\tau_l)\triangleq [\exp(-j2\pi\tau_l
f_s(1/\bar{K}))\phantom{0}\ldots\phantom{0}\exp(-j2\pi\tau_l f_s
(K/\bar{K}))]^T \label{g-tau-l}
\end{align}

Due to the sparse scattering nature of the mmWave channel, the
number of paths, $L$, is usually small relative to the dimensions
of the tensor. Hence the tensor $\boldsymbol{\mathcal{Y}}$ has an
intrinsic low-rank structure. As will be discussed later, this
low-rank structure ensures that the CP decomposition of
$\boldsymbol{\mathcal{Y}}$ is unique up to scaling and permutation
ambiguities. Therefore an estimate of the parameters
$\{\alpha_l,\phi_l,\theta_l,\tau_l\}$ as well as the mmWave
channels $\{\boldsymbol{H}_k\}$ can be obtained by performing a CP
decomposition of the received signal $\boldsymbol{\mathcal{Y}}$.
Define
\begin{align}
\boldsymbol{A}\triangleq &
[\boldsymbol{\tilde{a}}_{\text{MS}}(\theta_1)\phantom{0}\ldots\phantom{0}
\boldsymbol{\tilde{a}}_{\text{MS}}(\theta_L)] \label{factor-matrix-A}\\
\boldsymbol{B}\triangleq &
[\boldsymbol{\tilde{a}}_{\text{BS}}(\phi_1)\phantom{0}\ldots\phantom{0}
\boldsymbol{\tilde{a}}_{\text{BS}}(\phi_L)] \label{factor-matrix-B}\\
\boldsymbol{C}\triangleq &
[\boldsymbol{c}_1\phantom{0}\ldots\phantom{0}\boldsymbol{c}_L]
\label{factor-matrix-C}
\end{align}
These three matrices
$\{\boldsymbol{A},\boldsymbol{B},\boldsymbol{C}\}$ are factor
matrices associated with a noiseless version of
$\boldsymbol{\mathcal{Y}}$.

\subsection{CP Decomposition}
If the number of paths, $L$, is known or estimated \emph{a
priori}, the CP decomposition of $\boldsymbol{\mathcal{Y}}$ can be
accomplished by solving
\begin{align}
\min_{\boldsymbol{A},\boldsymbol{B},\boldsymbol{C}}\quad
\|\boldsymbol{\mathcal{Y}}-\sum_{l=1}^L
\boldsymbol{\tilde{a}}_{\text{MS}}(\theta_l)
\circ\boldsymbol{\tilde{a}}_{\text{BS}}(\phi_l)\circ
\boldsymbol{c}_l\|_{F}^2
\end{align}
The above optimization can be efficiently solved by an alternating
least squares (ALS) procedure which alternatively minimizes the
data fitting error with respect to one of the factor matrices,
with the other two factor matrices fixed
\begin{align}
\boldsymbol{A}^{(t+1)}=&\arg\min_{\boldsymbol{A}}
\left\|\boldsymbol{Y}_{(1)}^T-(\boldsymbol{C}^{(t)}\odot\boldsymbol{B}^{(t)})\boldsymbol{A}^T
\right\|_F^2 \label{A-update} \\
\boldsymbol{B}^{(t+1)}=&\arg\min_{\boldsymbol{B}}
\left\|\boldsymbol{Y}_{(2)}^T-(\boldsymbol{C}^{(t)}\odot\boldsymbol{A}^{(t+1)})\boldsymbol{B}^T
\right\|_F^2 \label{B-update} \\
{\boldsymbol{C}}^{(t+1)}=&\arg\min_{{\boldsymbol{C}}}
\left\|\boldsymbol{Y}_{(3)}^T-(\boldsymbol{B}^{(t+1)}
\odot\boldsymbol{A}^{(t+1)})\boldsymbol{C}^T \right\|_F^2
\label{C-update}
\end{align}


If the knowledge of the number of paths $L$ is unavailable, more
sophisticated CP decomposition techniques (e.g.
\cite{BazerqueMateos13,RaiWang14,ZhaoZhang15}) can be employed to
estimate the model order and the factor matrices simultaneously.
The basic idea of these CP decomposition techniques is to use
sparsity-promoting priors or functions to find a low-rank
representation of the observed tensor. As shown in
\cite{BazerqueMateos13}, the CP decomposition can still be solved
by an alternating least squares procedure as follows
\begin{align}
{{\boldsymbol{A}}^{(t+1)}} &= \arg\min_{\boldsymbol{A}} \left\|
{\left[ {\begin{array}{*{20}{c}}
{\boldsymbol{Y}_{(1)}^T}\\
\boldsymbol{0}
\end{array}} \right] - \left[ {\begin{array}{*{20}{c}}
{{{\boldsymbol{C}}^{(t)}} \odot {{\boldsymbol{B}}^{(t)}}}\\
{\sqrt \mu  {\boldsymbol{I}}}
\end{array}} \right]{{\boldsymbol{A}}^T}} \right\|_F^2 \nonumber\\
{{\boldsymbol{B}}^{(t + 1)}} &= \arg\min_{\boldsymbol{B}} \left\|
{\left[ {\begin{array}{*{20}{c}}
{\boldsymbol{Y}_{(2)}^T}\\
\boldsymbol{0}
\end{array}} \right] - \left[ {\begin{array}{*{20}{c}}
{{{\boldsymbol{C}}^{(t)}} \odot {{\boldsymbol{A}}^{(t+1)}}}\\
{\sqrt \mu  {\boldsymbol{I}}}
\end{array}} \right]{{\boldsymbol{B}}^T}} \right\|_F^2 \nonumber\\
{{\boldsymbol{C}}^{(t + 1)}} &= \arg\min_{\boldsymbol{C}} \left\|
{\left[ {\begin{array}{*{20}{c}}
{\boldsymbol{Y}_{(3)}^T}\\
\boldsymbol{0}
\end{array}} \right] - \left[ {\begin{array}{*{20}{c}}
{{{\boldsymbol{B}}^{(t+1)}} \odot {{\boldsymbol{A}}^{(t+1)}}}\\
{\sqrt \mu  {\boldsymbol{I}}}
\end{array}} \right]{{\boldsymbol{C}}^T}} \right\|_F^2 \nonumber
\end{align}
where $\boldsymbol{A}\in\mathbb{C}^{M\times \hat{L}}$,
$\boldsymbol{B}\in\mathbb{C}^{T\times \hat{L}}$,
$\boldsymbol{C}\in\mathbb{C}^{K\times \hat{L}}$, $\hat{L}>L$ is an
overestimated CP rank, and $\mu$ is a regularization parameter to
control the tradeoff between low-rankness and the data fitting
error. The true CP rank of the tensor, $L$, can be estimated by
removing those negligible rank-one tensor components after
convergence.

\subsection{Channel Estimation}
We discuss how to estimate the mmWave channels based on the
estimated factor matrices
$\{\boldsymbol{\hat{A}},\boldsymbol{\hat{B}},\boldsymbol{\hat{C}}\}$.
As shown in the next subsection, the CP decomposition is unique up
to scaling and permutation ambiguities under a mild condition.
More precisely, the estimated factor matrices and the true factor
matrices are related as
\begin{align}
\boldsymbol{\hat{A}}=&\boldsymbol{A}\boldsymbol{\Lambda}_1\boldsymbol{\Pi}+\boldsymbol{E}_1
\label{hat-Abs}
\\
\boldsymbol{\hat{B}}=&\boldsymbol{B}\boldsymbol{\Lambda}_2\boldsymbol{\Pi}+\boldsymbol{E}_2
\label{hat-Ams}
\\
\boldsymbol{\hat{C}}=&\boldsymbol{C}\boldsymbol{\Lambda}_3\boldsymbol{\Pi}+\boldsymbol{E}_3
\label{hat-F}
\end{align}
where
$\{\boldsymbol{\Lambda}_1,\boldsymbol{\Lambda}_2,\boldsymbol{\Lambda}_3\}$
are unknown nonsingular diagonal matrices which satisfy
$\boldsymbol{\Lambda}_1\boldsymbol{\Lambda}_2\boldsymbol{\Lambda}_3=\boldsymbol{I}$;
$\boldsymbol{\Pi}$ is an unknown permutation matrix; and
$\boldsymbol{E}_1$, $\boldsymbol{E}_2$, and $\boldsymbol{E}_3$
denote the estimation errors associated with the three estimated
factor matrices, respectively. The permutation matrix
$\boldsymbol{\Pi}$ can be ignored because it is common to all
three factor matrices. Note that each column of $\boldsymbol{A}$
is characterized by the associated angle of arrival $\theta_l$.
Hence the angle of arrival $\theta_l$ can be estimated via a
simple correlation-based method
\begin{align}
\hat{\theta}_l=\arg\max_{\theta_l}\quad
\frac{|\boldsymbol{\hat{a}}^H_l\boldsymbol{\tilde{a}}_{\text{MS}}(\theta_l)|}
{\|\boldsymbol{\hat{a}}_l\|_2\|\boldsymbol{\tilde{a}}_{\text{MS}}(\theta_l)\|_2}
\end{align}
where $\boldsymbol{\hat{a}}_l$ denotes the $l$th column of
$\boldsymbol{\hat{A}}$. It can be shown in Appendix \ref{appA}
that this simple correlation-based scheme is a maximum likelihood
(ML) estimator, provided that entries in the estimation error
matrix, $\boldsymbol{E}_1$, follow an i.i.d. circularly symmetric
Gaussian distribution. The angle of departure $\phi_l$ can be
obtained similarly as
\begin{align}
\hat{\phi}_l=\arg\max_{\phi_l}\quad
\frac{|\boldsymbol{\hat{b}}^H_l\boldsymbol{\tilde{a}}_{\text{BS}}(\phi_l)|}
{\|\boldsymbol{\hat{b}}_l\|_2\|\boldsymbol{\tilde{a}}_{\text{BS}}(\phi_l)\|_2}
\end{align}
where $\boldsymbol{\hat{b}}_l$ denotes the $l$th column of
$\boldsymbol{\hat{B}}$. We now discuss how to estimate the time
delay $\tau_l$ from the estimated factor matrix
$\boldsymbol{\hat{C}}$. Note that
$\boldsymbol{c}_l=\alpha_l\boldsymbol{g}(\tau_l)$. Therefore the
time delay $\tau_l$ can be estimated via
\begin{align}
\hat{\tau}_l=\arg\min_{\tau_l}\quad
\frac{|\boldsymbol{\hat{c}}_l^H\boldsymbol{g}(\tau_l)|}{\|\boldsymbol{\hat{c}}_l\|_2\|\boldsymbol{g}(\tau_l)\|_2}
\end{align}
where $\boldsymbol{\hat{c}}_l$ denotes the $l$th column of
$\boldsymbol{\hat{C}}$. Substituting the estimated $\{\theta_l\}$
and $\{\phi_l\}$ back into (\ref{hat-Abs}) and (\ref{hat-Ams}), an
estimate of the nonsingular diagonal matrices
$\boldsymbol{\Lambda}_1$ and $\boldsymbol{\Lambda}_2$ can be
obtained. An estimate of $\boldsymbol{\Lambda}_3$ can then be
calculated from the equality
$\boldsymbol{\Lambda}_1\boldsymbol{\Lambda}_2\boldsymbol{\Lambda}_3=\boldsymbol{I}$.
Finally, the fading coefficients $\{\alpha_l\}$ can be estimated
from (\ref{hat-F}). The channel matrices $\{\boldsymbol{H}_k\}$
can now be recovered from the estimated parameters
$\{\hat{\theta}_l,\hat{\phi}_l,\hat{\tau}_l,\hat{\alpha}_l\}$.


\subsection{Uniqueness}
We discuss the uniqueness of the CP decomposition. It is well
known that the essential uniqueness of CP decomposition can be
guaranteed by Kruskal's condition \cite{Kruskal77}. Let
$k_{\boldsymbol{X}}$ denote the k-rank of a matrix
$\boldsymbol{X}$, which is defined as the largest value of
$k_{\boldsymbol{X}}$ such that every subset of
$k_{\boldsymbol{X}}$ columns of the matrix $\boldsymbol{X}$ is
linearly independent. We have the following theorem.

\newtheorem{theorem}{Theorem}
\begin{theorem}
Let $(\boldsymbol{X},\boldsymbol{Y},\boldsymbol{Z})$ be a CP
solution which decomposes a third-order tensor
$\boldsymbol{\mathcal{X}}\in\mathbb{C}^{M\times N\times K}$ into
$R$ rank-one arrays, where $\boldsymbol{X}\in\mathbb{C}^{M\times
R}$, $\boldsymbol{Y}\in\mathbb{C}^{N\times R}$, and
$\boldsymbol{Z}\in\mathbb{C}^{K\times R}$. Suppose the following
Kruskal's condition
\begin{align}
k_{\boldsymbol{X}}+k_{\boldsymbol{Y}}+k_{\boldsymbol{Z}}\geq 2R+2
\label{Kruskals-condition}
\end{align}
holds and there is an alternative CP solution
$(\boldsymbol{\bar{X}},\boldsymbol{\bar{Y}},\boldsymbol{\bar{Z}})$
which also decomposes $\boldsymbol{\mathcal{X}}$ into $R$ rank-one
arrays. Then we have
$\boldsymbol{\bar{X}}=\boldsymbol{X}\boldsymbol{\Pi}\boldsymbol{\Lambda}_a$,
$\boldsymbol{\bar{Y}}=\boldsymbol{Y}\boldsymbol{\Pi}\boldsymbol{\Lambda}_b$,
and
$\boldsymbol{\bar{Z}}=\boldsymbol{Z}\boldsymbol{\Pi}\boldsymbol{\Lambda}_c$,
where $\boldsymbol{\Pi}$ is a unique permutation matrix and
$\boldsymbol{\Lambda}_a$, $\boldsymbol{\Lambda}_b$, and
$\boldsymbol{\Lambda}_c$ are unique diagonal matrices such that
$\boldsymbol{\Lambda}_a\boldsymbol{\Lambda}_b\boldsymbol{\Lambda}_c=\boldsymbol{I}$.
\end{theorem}
\begin{IEEEproof}
A rigorous proof can be found in \cite{StegemanSidiropoulos07}.
\end{IEEEproof}
Note that Kruskal's condition is necessary and sufficient for
uniqueness when $R\geq 2$, but it is not necessary for $R=1$.


From the above theorem, we know that if
\begin{align}
k_{{{\boldsymbol{A}}}}+k_{\boldsymbol{B}}+k_{\boldsymbol{C}}\geq
2L+2 \label{Kruskals-condition-v2}
\end{align}
then the CP decomposition of $\boldsymbol{\mathcal{Y}}$ is
essentially unique.

We first examine the k-rank of $\boldsymbol{A}$. Note that
\begin{align}
\boldsymbol{A}=\boldsymbol{Q}^T[\boldsymbol{a}_{\text{MS}}(\theta_1)\phantom{0}\ldots\phantom{0}
\boldsymbol{a}_{\text{MS}}(\theta_L)]\triangleq\boldsymbol{Q}^T\boldsymbol{A}_{\text{MS}}
\end{align}
where
$\boldsymbol{A}_{\text{MS}}\in\mathbb{C}^{N_{\text{MS}}\times L}$
is a Vandermonte matrix when a uniform linear array is employed.
Suppose assumption A1 holds valid. For a randomly generated
$\boldsymbol{Q}$ whose entries are chosen uniformly from a unit
circle, we can show that the k-rank of $\boldsymbol{A}$ is equal
to (details can be found in Appendix \ref{appB})
\begin{align}
k_{{{\boldsymbol{A}}}}=\min(M,L) \label{k-rank-A}
\end{align}
with probability one. Similarly, for a randomly generated
$\boldsymbol{P}$ whose entries are uniformly chosen from a unit
circle, we can deduce that the k-rank of $\boldsymbol{B}$ is equal
to
\begin{align}
k_{{{\boldsymbol{B}}}}=\min(T,L) \label{k-rank-B}
\end{align}
with probability one. Now let us examine the k-rank of
$\boldsymbol{C}$. Recall that $\boldsymbol{C}$ can be expressed as
\begin{align}
\boldsymbol{C}=[\boldsymbol{g}(\tau_1)\phantom{0}\ldots\phantom{0}\boldsymbol{g}(\tau_L)]\boldsymbol{D}_{\alpha}
\end{align}
where
$\boldsymbol{D}_{\alpha}\triangleq\text{diag}(\alpha_1,\ldots,\alpha_L)$,
and $\boldsymbol{g}(\tau_l)$ is defined in (\ref{g-tau-l}). We see
that $\boldsymbol{C}$ is a columnwise-scaled Vandermonte matrix.
Therefore the k-rank of $\boldsymbol{C}$ is
\begin{align}
k_{{{\boldsymbol{C}}}}=\min(K,L)
\end{align}


Since $L$ is usually small, it is reasonable to assume that the
number of subcarriers used for training is greater than $L$, i.e.
$K\geq L$. Hence we have $k_{\boldsymbol{C}}=L$. To meet Kruskal's
condition (\ref{Kruskals-condition-v2}), we only need
$k_{\boldsymbol{A}}+k_{\boldsymbol{B}}\geq L+2$. Recalling
(\ref{k-rank-A})--(\ref{k-rank-B}), we can either choose
$\{T=L,M=2\}$ or $\{M=L,T=2\}$ to satisfy Kruskal's condition. In
summary, for randomly generated beamforming matrix
$\boldsymbol{P}$ and combining matrix $\boldsymbol{Q}$ whose
entries are chosen uniformly from a unit circle, our proposed
method only needs $T=L$ (or $T=2$) time frames and $M=2$ (or
$M=L$) sub-frames to enable reliable estimation of channel
parameters, thus achieving a substantial training overhead
reduction. In practice, due to the observation noise and
estimation errors, we may need a slightly larger $T$ and $M$ to
yield an accurate channel estimate. Note that besides random
coding, coded beams \cite{TsangPoon11} which steer the antenna
array towards multiple beam directions simultaneously can also be
used to serve as the beamforming and combining vectors
$\{\boldsymbol{p}_t\}$ and $\{\boldsymbol{q}_m\}$. The k-ranks of
$\boldsymbol{A}$ and $\boldsymbol{B}$ may still obey
(\ref{k-rank-A}) and (\ref{k-rank-B}) if the coded beams are
carefully designed. The design of the coded beams for our proposed
method will be explored in our future work.

\section{CRB} \label{sec:CRB}
In this section, we develop Cram\'{e}r-Rao bound (CRB) results for
the channel parameter (i.e.
$\{\hat{\theta}_l,\hat{\phi}_l,\hat{\tau}_l,\hat{\alpha}_l\}$)
estimation problem considered in (\ref{Y-tensor}). Details of the
derivation can be found in Appendix \ref{appC}. Throughout our
analysis, the observation noise in (\ref{Y-tensor}) is assumed to
be complex circularly symmetric i.i.d. Gaussian noise. As is well
known, the CRB is a lower bound on the variance of any unbiased
estimator \cite{Kay93}. It provides a benchmark for evaluating the
performance of our proposed method. In addition, the CRB results
illustrate the behavior of the resulting bounds, which helps
understand the effect of different system parameters, including
the beamforming and combining matrices, on the estimation
performance.

Note that our proposed method involves two steps: the first step
employs an ALS algorithm to perform the CP decomposition, and
based on the decomposed factor matrices, the second step uses a
simple correlation-based method to estimate the channel
parameters. For zero-mean i.i.d. Gaussian noise, the ALS yields
maximum likelihood estimates \cite{LiuSidiropoulos01}, provided
that the global minimum is reached. Also, it can be proved that
the correlation-based method used in the second step is a maximum
likelihood estimator if the estimation errors associated with the
factor matrices are i.i.d. Gaussian random variables. Therefore
our proposed method can be deemed as a quasi-maximum likelihood
estimator for the channel parameters. Under mild regularity
conditions, the maximum likelihood estimator is asymptotically (in
terms of the sample size) unbiased and asymptotically achieves the
CRB. It therefore makes sense to compare our proposed
CP-decomposition-based method with the CRB results.




\section{Compressed Sensing-Based Channel Estimation}\label{sec:cs-method}
By exploiting the sparse scattering nature, the downlink channel
estimation problem considered in this paper can also be formulated
as a sparse signal recovery problem. In the following, we briefly
discuss this compressed sensing-based channel estimation method.

Taking the mode-3 unfolding of $\boldsymbol{\mathcal{Y}}$ (c.f.
(\ref{Y-tensor})), we have
\begin{align}
\boldsymbol{Y}_{(3)} &= {\boldsymbol{C}}{({{\boldsymbol{B}}} \odot
{{\boldsymbol{A}}})^T}+ \boldsymbol{W}_{(3)}
\nonumber\\
&= {\boldsymbol{G}\boldsymbol{D}_{\alpha} }{{\boldsymbol{\Sigma
}}^T}{( {\boldsymbol{P}}^T\otimes{\boldsymbol{Q}}^T
)^T}+\boldsymbol{W}_{(3)}
\end{align}
where
\begin{align}
{\boldsymbol
G}\triangleq&[\boldsymbol{g}(\tau_1)\phantom{0}\ldots\phantom{0}\boldsymbol{g}(\tau_L)],
\quad
\boldsymbol{D}_{\alpha}\triangleq\text{diag}(\alpha_1,\ldots,\alpha_L),
\nonumber\\
{\boldsymbol{\Sigma }} \triangleq &[\boldsymbol{a}_{\text
{BS}}(\phi_1)\otimes\boldsymbol{a}_{\text{MS}}(\theta_1)\phantom{0}
\cdots \phantom{0} \boldsymbol{a}_{\text
{BS}}(\phi_L)\otimes\boldsymbol{a}_{\text{MS}}(\theta_L)]
\nonumber
\end{align}
Taking the transpose of $\boldsymbol{Y}_{(3)}$, we have
\begin{align}
{\boldsymbol{Y}}_{(3)}^T =
({\boldsymbol{P}}^T\otimes{\boldsymbol{Q}}^T){\boldsymbol{\Sigma
}}\boldsymbol{D}_{\alpha}{{\boldsymbol{G}}^T} +
\boldsymbol{W}_{(3)} \label{mode-3-unfolding-Y}
\end{align}
We see that both $\boldsymbol{\Sigma}$ and $\boldsymbol{G}$ are
characterized by unknown parameters which need to be estimated. To
convert the estimation problem into a sparse signal recovery
problem, we discretize the AoA-AoD space into an $N_1\times N_2$
grid, in which each grid point is given by $\{{\bar
\theta_i},{\bar \phi_j}\}$ for $i=1,...,N_1$ and $j=1,...,N_2$,
where $N_1\gg L$, and $N_2\gg L$. The true angles of
arrival/departure are assumed to lie on the grid. Also, we
discretize the time-delay domain into a finite set of grid points
$\{\bar{\tau}_l\}_{l=1}^{N_3}$ ($N_3\gg L$), and assume that the
true time-delays $\{\tau_l\}$ lie on the discretized grid. Thus
(\ref{mode-3-unfolding-Y}) can be re-expressed as
\begin{align}
\boldsymbol{Y}_{(3)}^T =
({\boldsymbol{P}}^T\otimes{\boldsymbol{Q}}^T)\boldsymbol{\bar
\Sigma} \boldsymbol{\bar{D}}_{\alpha}\boldsymbol{ \bar G}^T +
\boldsymbol{W}_{(3)} \label{mode-3-unfolding-Y-new}
\end{align}
where
$\boldsymbol{\bar{\Sigma}}\in\mathbb{C}^{N_{\text{MS}}N_{\text{BS}}\times
N_{1}N_2}$ is an overcomplete dictionary consisting of $N_1 N_2$
columns, with its $(i+(j-1)N_1)$th column given by
$\boldsymbol{a}_{\text{BS}}(\bar{\phi_j})\otimes\boldsymbol{a}_{\text{MS}}(\bar{\theta_i})$,
and $\boldsymbol{\bar G}\in\mathbb{C}^{K\times N_3}$ is an
overcomplete dictionary, with its $n$th column given by
$\boldsymbol{g}(\bar{\tau}_n)$. $\boldsymbol{\bar{D}}_{\alpha}$ is
a sparse matrix obtained by augmenting ${\boldsymbol D_{\alpha}}$
with zero rows and columns. Let
${\boldsymbol{y}}\triangleq\text{vec}({{\boldsymbol{Y}}_{(3)}^T})$,
(\ref{mode-3-unfolding-Y-new}) can be formulated as a conventional
sparse signal recovery problem
\begin{equation}
{\boldsymbol{y}} = {\boldsymbol{\bar G}} \otimes
(({\boldsymbol{P}}^T\otimes{\boldsymbol{Q}}^T){\boldsymbol{\bar
\Sigma }}){\boldsymbol{d}} +\boldsymbol{w} \label{eqn1}
\end{equation}
where
${\boldsymbol{d}}\triangleq\text{vec}(\boldsymbol{\bar{D}}_{\alpha})$
is an unknown sparse vector. Many efficient algorithms such as the
orthogonal matching pursuit (OMP) \cite{PatiRezaiifar93} or the
fast iterative shrinkage-thresholding algorithm (FISTA)
\cite{BeckTeboulle09} can be employed to solve the above sparse
signal recovery problem. In practice, the true parameters do not
necessarily lie on the discretized grid. This error, also referred
to as the grid mismatch, leads to deteriorated performance. To
address this issue, one can employ finer grids to reduce the grid
mismatch error. Nevertheless, a finer grid not only results in a
higher computational complexity, but also brings the issue of
numerical instability due to the high coherence between columns of
the dictionary. Another solution is to employ super-resolution
(also referred to as off-grid) compressed sensing techniques (e.g.
\cite{HuShi12,YangXie13,FangWang16}) to mitigate the
discretization errors. This class of approaches have a high
computational complexity because they usually involve an iterative
process for joint dictionary refinement and sparse signal
estimation.




\section{Computational Complexity Analysis} \label{sec:complexity-analysis}
We analyze the computational complexity of the proposed CP
decomposition-based method and the compressed sensing method
discussed in the previous section. The major computational task of
our proposed method involves solving the three least squares
problems (\ref{A-update})--(\ref{C-update}) at each iteration.
Considering the calculation of $\boldsymbol A$, we have
\begin{align}
\boldsymbol{A}^{(t+1)}=
{{\boldsymbol{Y}}_{(1)}}{{\boldsymbol{V}}^*}(
\boldsymbol{V}^{T}\boldsymbol{V}^{*})^{-1}
\end{align}
where $\boldsymbol{V}\triangleq {{{\boldsymbol{C}}}^{(t)}} \odot
{\boldsymbol{B}}^{(t)}\in {\mathbb C}^{T K\times L}$ is a tall
matrix since we usually have $T K>L$. Noting that
$\boldsymbol{Y}_{(1)}\in M\times TK$, the number of flops required
to compute ${\boldsymbol A}^{(t+1)}$ is of order ${\cal O}(MTKL +
M K{L^2} + {L^3})$. When $L$ is small, the dominant term has a
computational complexity of order ${\cal O}(MTK)$, which scales
linearly with the size of the observed tensor $\boldsymbol {\cal
Y}$. It can also be shown that solving the other two least squares
problems requires flops of order ${\cal O}(MTK)$ as well.


The compressed sensing method discussed in the previous section
involves finding a sparse solution to the linear equation
(\ref{eqn1}). As indicated earlier, many efficient compressed
sensing algorithms such as greedy methods (e.g.
\cite{PatiRezaiifar93}) or $\ell_1$-minimization-based methods
(e.g. \cite{BeckTeboulle09}) can be employed to solve
(\ref{eqn1}). Greedy methods such as the orthogonal matching
pursuit (OMP) have a low computational complexity but usually
yield barely satisfactory recovery accuracy. In contrast,
$\ell_1$-minimization-based methods achieve better performance but
incur higher computational complexity. It can be easily verified
that the computational complexity of the OMP is of order ${\cal
O}(MTK+N_1 N_2 N_3)$. For the FISTA \cite{BeckTeboulle09}, the
main computational task at each iteration is to evaluate the
proximal operator whose computational complexity is of the order
${\cal O}(n^2)$, where $n$ denotes the number of columns of the
overcomplete dictionary. For our case, we have $n=N_1 N_2 N_3$.
Thus the required number of flops at each iteration is of order
${\cal O}(N_1^{2}N_{2}^{2}N_3^2)$, which scales quadratically with
$N_1N_2N_3$. In order to achieve a substantial overhead reduction,
the parameters $\{M,T,K\}$ are usually chosen such that the number
of measurements is far less than the dimension of the sparse
signal, i.e., $MTK \ll {N_1}{N_2}N_3$. Therefore we see that both
the OMP and the FISTA have a higher computational complexity than
our proposed CP decomposition-based method.



\section{Simulation Results} \label{sec:experiments}
We present simulation results to illustrate the performance of our
proposed CP decomposition-based method (referred to as CP). We
consider a scenario where the BS employs a uniform linear array
with $N_{\text{BS}}=64$ antennas and the MS employs a uniform
linear array with $N_{\text{MS}}=32$ antennas. The distance
between neighboring antenna elements is assumed to be half the
wavelength of the signal. In our simulations, the mmWave channel
is generated according to the wideband geometric channel model, in
which the AoAs and AoDs are randomly distributed in $[0,2\pi]$,
the number of paths is set equal to $L=4$, the delay spread
$\tau_l$ for each path is uniformly distributed between $0$ and
$100$ nanoseconds, and the complex gain $\alpha_l$ is a random
variable following a circularly-symmetric Gaussian distribution
$\alpha_{u,l}\sim\mathcal{CN}(0,1/\rho)$. Here $\rho$ is given by
$\rho=(4\pi D f_c/c)^2$, where $c$ represents the speed of light,
$D$ denotes the distance between the MS and the BS, and $f_c$ is
the carrier frequency. We set $D=100$m, $f_c=28$GHz. The total
number of subcarriers is set to $\bar{K}=128$, out of which $K$
subcarriers are selected for training. The sampling rate is set to
$f_s=0.32$GHz. Also, in our experiments, the beamforming matrix
$\boldsymbol{P}$ and the combining matrix $\boldsymbol{Q}$ are
randomly generated with their entries uniformly chosen from a unit
circle. The signal-to-noise ratio (SNR) is defined as the ratio of
the signal component to the noise component, i.e.
\begin{align}
\text{SNR}\triangleq
\frac{\|\boldsymbol{\mathcal{Y}}-\boldsymbol{\mathcal{W}}\|_F^2}{\|\boldsymbol{\mathcal{W}}\|_F^2}
\end{align}
where $\boldsymbol{\mathcal{Y}}$ and $\boldsymbol{\mathcal{W}}$
represent the received signal and the additive noise in
(\ref{Y-tensor}), respectively.

\begin{figure}[!t]
\centering
\includegraphics[width=8.5cm]{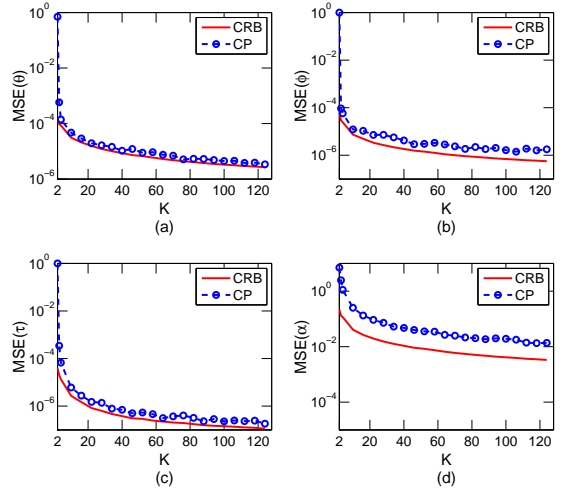}
\caption{MSEs and CRBs associated with different sets of
parameters vs. the number of subcarriers, $K$.} \label{fig2}
\end{figure}

\begin{figure}[!t]
\centering
\includegraphics[width=8.5cm]{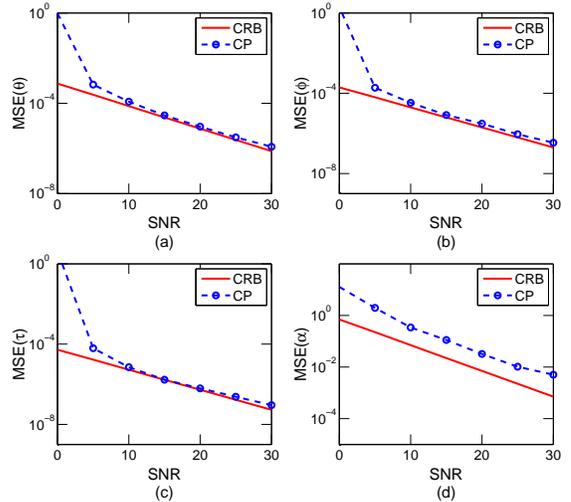}
\caption{MSEs and CRBs associated with different sets of
parameters vs. SNR.} \label{fig3}
\end{figure}

We first examine the estimation accuracy of the channel parameters
$\{\theta_l,\phi_l,\tau_l,\alpha_l\}$. Mean square errors (MSEs)
are calculated separately for each set of parameters, i.e.
\begin{align}
\text{MSE}(\theta)=&\|\boldsymbol{\theta}-\boldsymbol{\hat{\theta}}\|_2^2
\quad
\text{MSE}(\phi)=\|\boldsymbol{\phi}-\boldsymbol{\hat{\phi}}\|_2^2 \nonumber\\
\text{MSE}(\tau)=&\|\boldsymbol{\tau}-\boldsymbol{\hat{\tau}}\|_2^2
\quad
\text{MSE}(\alpha)=\|\boldsymbol{\alpha}-\boldsymbol{\hat{\alpha}}\|_2^2
\nonumber
\end{align}
where $\boldsymbol{\theta}\triangleq
[\theta_1\phantom{0}\ldots\phantom{0}\theta_L]^T$,
$\boldsymbol{\phi}\triangleq
[\phi_1\phantom{0}\ldots\phantom{0}\phi_L]^T$,
$\boldsymbol{\tau}\triangleq
[\tau_1\phantom{0}\ldots\phantom{0}\tau_L]^T$, and
$\boldsymbol{\alpha}\triangleq
[\alpha_1\phantom{0}\ldots\phantom{0}\alpha_L]^T$. Fig. \ref{fig2}
plots the MSEs of our proposed method as a function of the number
of subcarriers used for training, $K$, where we set $M=6$, $T=6$,
and $\text{SNR}=10\text{dB}$. The CRB results for different sets
of parameters are also included for comparison. We see that our
proposed method yields accurate estimates of the channel
parameters even for small values of $M$, $T$, and $K$. This result
indicates that our proposed method is able to achieve a
substantial training overhead reduction. We also notice that the
MSEs attained by our proposed method are very close to their
corresponding CRBs, particularly for the AoA, AoD, and the time
delay parameters. This result corroborates the optimality of the
proposed method. As indicated earlier, the optimality of the
proposed method comes from the fact that the ASL and the
correlation-based scheme used in our proposed method are all
maximum likelihood estimators under mild conditions. Specifically,
it has been shown in \cite{LiuSidiropoulos01} that the ALS yields
maximum likelihood estimates and achieves its associated CRB in
the presence of zero-mean i.i.d. Gaussian noise. We also proved
that the simple correlation-based scheme employed in the second
stage of our proposed method is a maximum likelihood estimator,
provided that the estimator errors of the factor matrices are
i.i.d. Gaussian random variables. Although the estimator errors
may not strictly follow an i.i.d. Gaussian distribution, the
correlation-based scheme is still an effective estimator that
achieves near-optimality. Lastly, we observe that our proposed
method fails when the number of subcarriers $K\leq 2$. This is
because, for the case where $T=6>L$ and $M=6>L$, Kruskal's
condition is satisfied only if $K\geq 2$. Thus our result roughly
coincides with our previous analysis regarding the uniqueness of
the CP decomposition. Fig. \ref{fig3} depicts the MSEs and CRBs
vs. SNR, where we set $T=6$, $M=6$, and $K=6$. From Fig.
\ref{fig3}, we see that the CRBs decrease exponentially with
increasing SNR, and the estimation accuracy achieved by our
proposed method has similar tendency as the CRBs. The MSEs
attained by our proposed method, again, are close to their
corresponding CRBs, except in the low SNR regime.






\begin{figure}[!t]
\centering
\includegraphics[width=8cm]{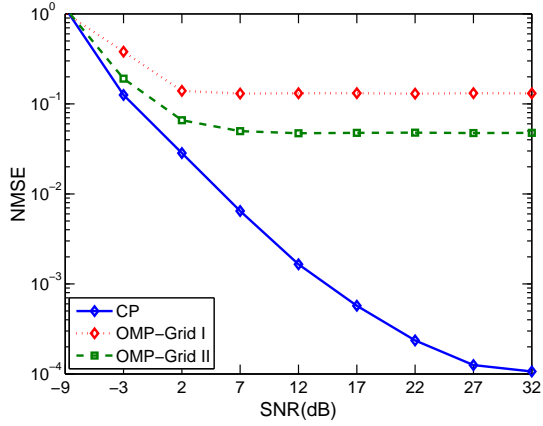}
\caption{NMSEs of respective algorithms vs. SNR, $M=6$, $T=6$,
$K=6$.} \label{fig4}
\end{figure}

\begin{figure}[!t]
\centering
\includegraphics[width=8cm]{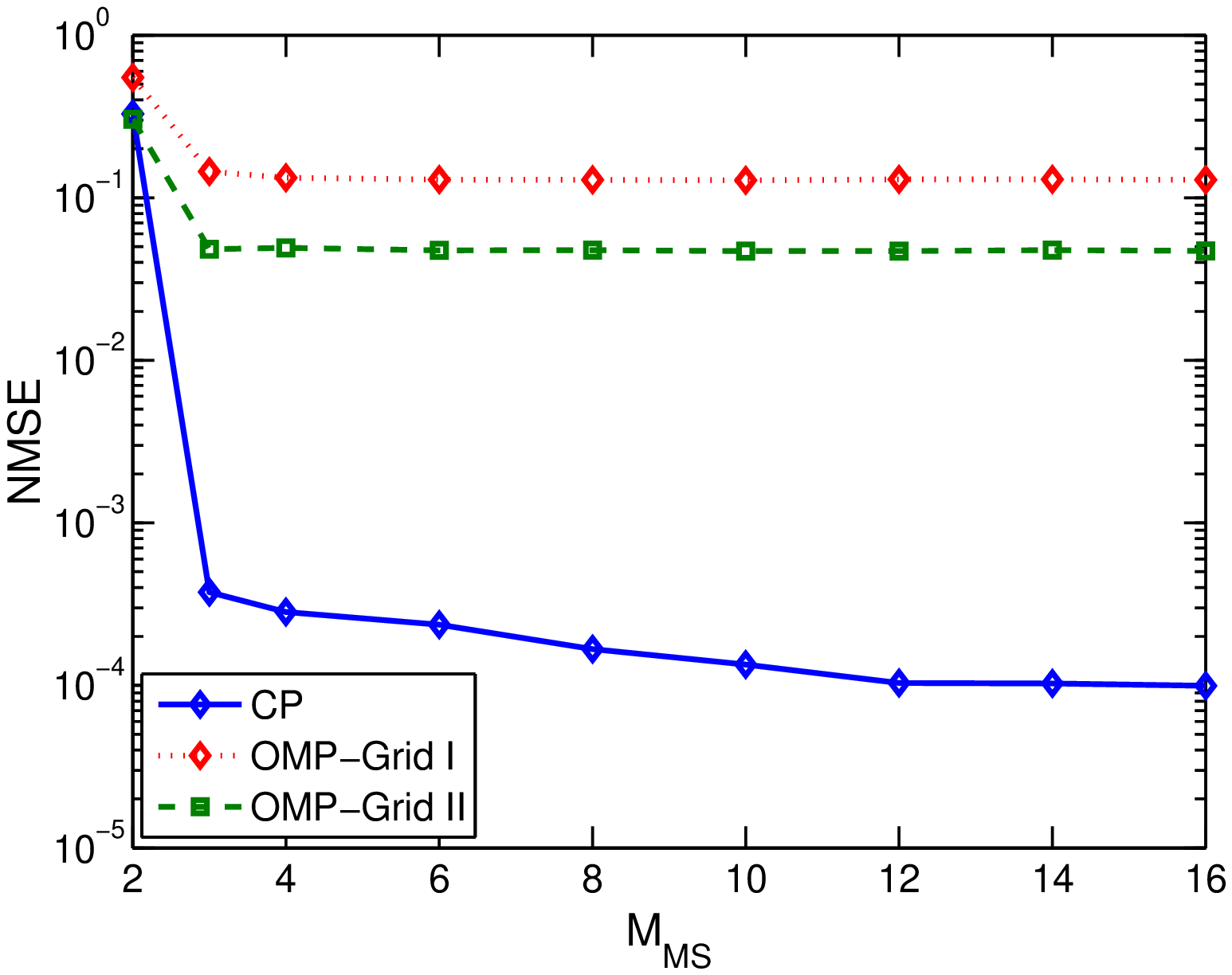}
\caption{NMSEs of respective algorithms vs. the number of
sub-frames $M$, $T=6$, $K=6$.} \label{fig5}
\end{figure}



\begin{figure}[!t]
\centering
\includegraphics[width=8cm]{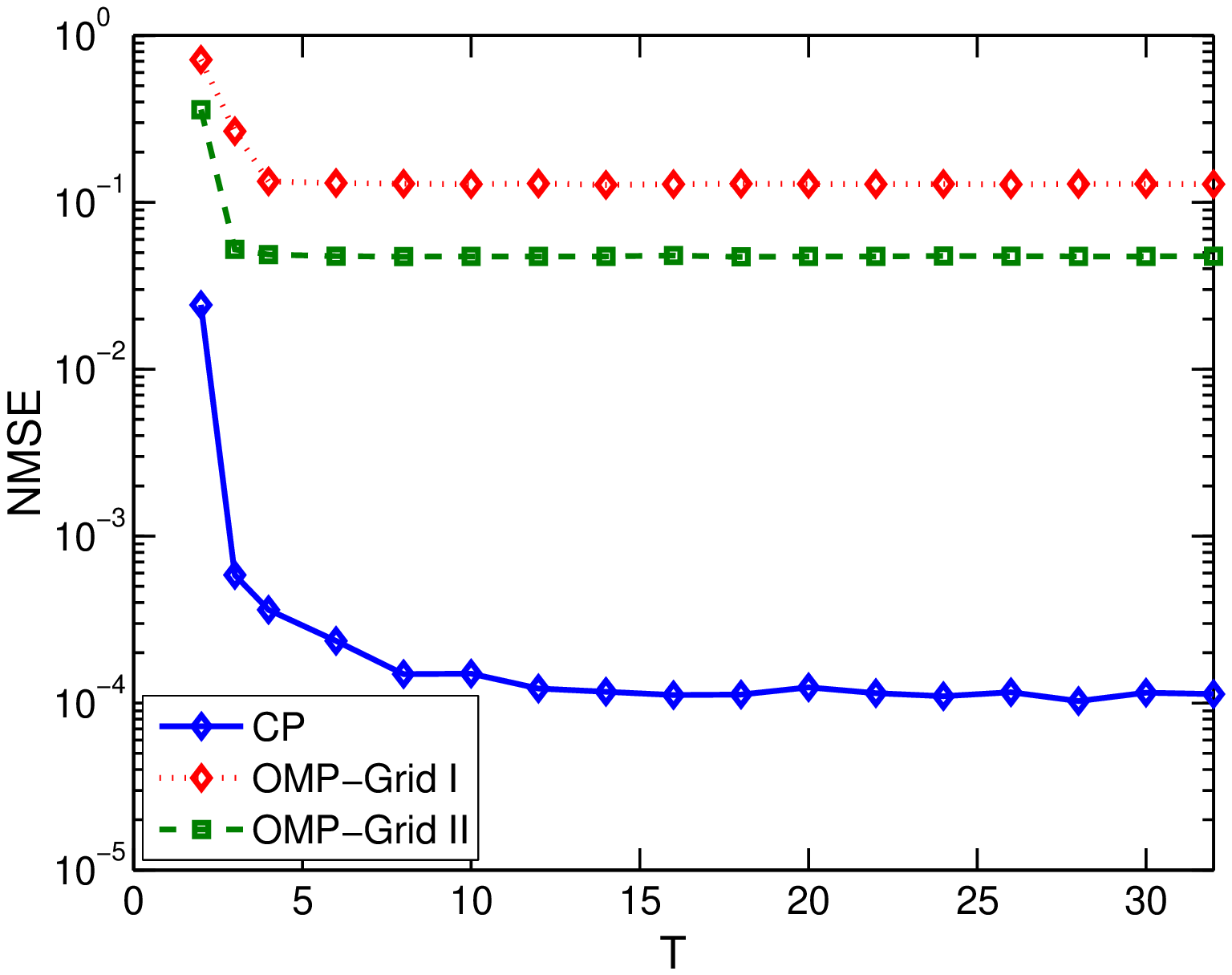}
\caption{NMSEs of respective algorithms vs. the number of frames
$T$, $M=6$, $K=6$.} \label{fig6}
\end{figure}

\begin{figure}[!t]
\centering
\includegraphics[width=8cm]{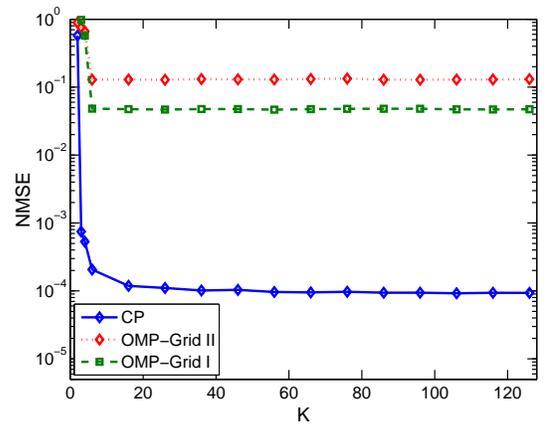}
\caption{NMSEs of respective algorithms vs. the number of
subcarriers for training $K$, $M=6$, $T=6$.} \label{fig7}
\end{figure}

We now examine the channel estimation performance of our proposed
method and its comparison with the compressed sensing method
discussed in Section \ref{sec:cs-method}. Specifically, an
orthogonal matching pursuit (OMP) algorithm is employed to solve
the sparse signal recovery problem (\ref{eqn1}). Note that the
dimension of the signal to be recovered in (\ref{eqn1}) is equal
to $N_1 N_2 N_3$, where $N_1$, $N_2$, and $N_3$ denote the number
of grid points used to discretize the AoA, AoD, and time delay
domain, respectively. For a typical choice of $N_1=32$, $N_2=64$
and $N_3=32$, the dimension of the signal is of order
$\mathcal{O}(10^4)$. In this case, more sophisticated sparse
recovery algorithms such as the fast iterative
shrinkage-thresholding algorithm (FISTA) have a prohibitive
computational complexity and thus are not included. Also, for the
OMP, we employ two different grids to discretize the continuous
parameter space: the first grid (referred to as Grid-I)
discretizes the AoA-AoD-time delay space into $64\times 128\times
256$ grid points, and the second grid (referred to as Grid-II)
discretizes the AoA-AoD-time delay space into $128\times 256\times
512$ grid points.

In Fig. \ref{fig4}, we show the NMSE results for our proposed
method and the OMP algorithm as a function of SNR, where we set
$M=6$, $T=6$, and $K=6$. Here the NMSE is calculated as
\begin{align}
\text{NMSE}=\frac{\sum_{k=1}^K\|\boldsymbol{H}_k-
\boldsymbol{\hat{H}}_k\|_F^2}{\sum_{k=1}^K\|\boldsymbol{H}_k\|_F^2}
\end{align}
where $\boldsymbol{H}_k$ denotes the frequency-domain channel
matrix associated with the $k$th subcarrier, and
$\boldsymbol{\hat{H}}_k$ is its estimate. We see that our proposed
method achieves a substantial performance improvement over the
compressed sensing algorithm. The performance gain is primarily
due to the following two reasons. First, unlike compressed sensing
techniques, our proposed CP decomposition-based method is
essentially a gridless approach which is free from grid
discretization errors. Second, the CP decomposition-based method
captures the intrinsic multi-dimensional structure of the multiway
data, which helps lead to a performance improvement. From Fig.
\ref{fig5} to Fig. \ref{fig7}, we plot the NMSEs of respective
methods vs. $M$, $T$, and $K$, respectively, where the SNR is set
to 20dB. These results, again, demonstrate the superiority of the
proposed method over the compressed sensing method. We also
observe that these results corroborate our theoretical analysis
concerning the uniqueness of the CP decomposition. For example, in
Fig. \ref{fig5}, since we have $T=6>L$ and $K=6>L$, we only need
$M\geq 2$ to satisfy Kruskal's condition. We see that our proposed
method achieves an accurate channel estimate only when $M>2$,
which roughly coincides with our analysis.

Table \ref{table1} shows the average run times of our proposed
method and the OMP method. To provide a glimpse of other more
sophisticated compressed sensing method's computational
complexity, the average run times of the FISTA are also included,
from which we can see that sophisticated compressed sensing
methods have a prohibitive computational complexity, and thus are
not suitable for our channel estimation problem. We also see that
our proposed method has a computational complexity as low as the
OMP method. It takes similar run times as the OMP method which
employs the coarser grid of the two choices, meanwhile achieving a
much higher estimation accuracy than the OMP method that uses the
finer grid.

\begin{table}[!t]
\caption{Average run times of respective algorithms: $T=6$, $M=6$,
$K=6$, and $\text{SNR}=20\text{dB}$.}
\renewcommand{\arraystretch}{1.3}
\centering
\begin{tabular}{|c|c|c|c|}
\hline
{\bfseries{ALG}}&{\bfseries{Grid}}&
{\bfseries{ NMSE}}&{\bfseries{ Average Run Time(s)}}\\
\hline
{\multirow{2}*{OMP}}&$64\times128\times256$&$2.3e-1$&$0.2$\\
\cline{2-4}
 &$160\times320\times640$&$3.5e-2$&$0.8$\\
\hline
{\multirow{2}*{FISTA}}&$32\times64\times128$&$2.3e-1$&$9e2$\\
\cline{2-4}
 &$64\times128\times256$&$9e-2$&$6e3$\\
 \hline
CP&-&$1.2e-4$&$0.2$\\
\hline
\end{tabular}
\label{table1}
\end{table}

\section{Conclusions} \label{sec:conclusion}
We proposed a CP decomposition-based method for downlink channel
estimation in mm-Wave MIMO-OFDM systems, where wideband mmWave
channels with frequency selectivity were considered. The proposed
method exploited the intrinsic multi-dimensional structure of the
multiway data received at the BS. Specifically, the received
signal at the BS was expressed as a third-order tensor. We showed
that the tensor has a form of a low-rank CP decomposition, and the
channel parameters can be easily extracted from the decomposed
factor matrices. The uniqueness of the CP decomposition was
investigated, which revealed that the uniqueness of the CP
decomposition can be guaranteed even with a small number of
measurements. Thus the proposed method is able to achieve a
substantial training overhead reduction. CRB results for channel
parameters were also developed. We compared our proposed method
with a compressed sensing-based channel estimation method.
Simulation results showed that our proposed method presents a
clear performance advantage over the compressed sensing method in
terms of both estimation accuracy and computational complexity.

\appendices

\section{} \label{appA}
In (\ref{hat-Abs}), for the $l$th column, we have
\begin{align}\label{observation_l}
\boldsymbol{\hat{a}}_l=\lambda_l\boldsymbol{\tilde{a}}_{\rm{MS}}(\theta_l)
+ \boldsymbol{e}_l
\end{align}
where $\theta_l$ and $\lambda_l$ are unknown parameters. We assume
$\boldsymbol{e}_l$ satisfies circularly symmetric complex Gaussian
distribution with zero mean and covariance matrix
$\epsilon^2\boldsymbol{I}$. Thus the log-likelihood function is
given by
\begin{align*}
L({\theta _l},{\lambda _l}) &=-{M_{\rm MS}}\ln (\pi {\epsilon ^2})
- \frac{1}{{{\epsilon ^2}}}\left\| {{{{\boldsymbol{\hat a}}}_l} -
{\lambda _l}{{{\boldsymbol{\tilde a}}}_{{\rm{MS}}}}({\theta _l})} \right\|_F^2\\
    &\propto  - \left\| {{{{\boldsymbol{\hat a}}}_l} -
    {\lambda _l}{{{\boldsymbol{\tilde a}}}_{{\rm{MS}}}}({\theta _l})} \right\|_2^2
\end{align*}
Given a fixed $\theta_l$, the optimal $\lambda_l$ can be obtained
by taking the partial derivative of the log-likelihood function
with respect to $\lambda_l$ and setting the partial derivative
equal to zero, i.e.
\[\frac{{\partial L({\theta _l},{\lambda _l})}}{{\partial {\lambda _l^*}}} =
{({{{\boldsymbol{\hat a}}}_l} - {\lambda _l}{{{\boldsymbol{\tilde
a}}}_{{\rm{MS}}}}({\theta _l}))^T} {{{\boldsymbol{\tilde
a}}}_{{\rm{MS}}}^*}({\theta _l}) = 0\]
which leads to
\[\lambda _l^\star = \frac{{{{{\boldsymbol{\hat a}}}_l}^T{{{\boldsymbol{\tilde a}}}_{{\rm{MS}}}^*}({\theta _l})}}
{{{{\left\| {{{{\boldsymbol{\tilde a}}}_{{\rm{MS}}}}({\theta _l})} \right\|}^2}}}\]
Note that (\ref{observation_l}) can be rewritten as
\[{\left\| {{{{\boldsymbol{\hat a}}}_l}} \right\|^2} =
{\lambda _l}{\boldsymbol{\hat a}}_l^H{{{\boldsymbol{\tilde
a}}}_{{\rm{MS}}}}({\theta _l}) + {\boldsymbol{\hat
a}}_l^H{{\boldsymbol{e}}_l}\] Then the log-likelihood function
becomes
\begin{align}
    L({\theta _l},{\lambda _l}) &\propto  - {\left\| {{{\left\| {{{{\boldsymbol{\hat a}}}_l}}
    \right\|}^2} - {\lambda _l}{\boldsymbol{\hat a}}_l^H{{{\boldsymbol{\tilde a}}}_{{\rm{MS}}}}({\theta _l})} \right\|^2}
\end{align}
Substituting $\lambda_l^{\star}$ into the above log-likelihood
function, we arrive at
\begin{align}
L({\theta _l},\lambda _l^ \star ) \propto  - \left\| {{{\left\| {{{{\boldsymbol{\hat a}}}_l}} \right\|}^2}
 - \frac{{{{\left| {{\boldsymbol{\hat a}}_l^H{{{\boldsymbol{\tilde a}}}_{{\rm{MS}}}}({\theta _l})}
 \right|}^2}}}{{{{\left\| {{{{\boldsymbol{\tilde a}}}_{{\rm{MS}}}}({\theta _l})} \right\|}^2}}}} \right\|^2\\
    \propto  - \left\| {1 - \frac{{{{\left| {{\boldsymbol{\hat a}}_l^H{{{\boldsymbol{\tilde a}}}_{{\rm{MS}}}}({\theta _l})}
    \right|}^2}}}{{{{\left\| {{{{\boldsymbol{\tilde a}}}_{{\rm{MS}}}}({\theta _l})} \right\|}^2}
    {{\left\| {{{{\boldsymbol{\hat a}}}_l}} \right\|}^2}}}}
    \right\|^2
\end{align}
Due to the Cauchy-Schwarz inequality, we have
\[0 \le \frac{{{{\left| {{\boldsymbol{\hat a}}_l^H{{{\boldsymbol{\tilde a}}}_{{\rm{MS}}}}({\theta _l})}
\right|}^2}}}{{{{\left\| {{{{\boldsymbol{\tilde
a}}}_{{\rm{MS}}}}({\theta _l})} \right\|}^2}{{\left\|
{{{{\boldsymbol{\hat a}}}_l}} \right\|}^2}}} \le 1\] Therefore
maximizing the log-likelihood with respect to $\theta_l$ is
equivalent to
\begin{align}
    \theta _l^\star  = \mathop {\arg \max }\limits_{{\theta _l}} L({\theta _l},\lambda _l^ \star )
    = \mathop {\arg \max }\limits_{{\theta _l}} \frac{{{{\left|
    {{\boldsymbol{\hat a}}_l^H{{{\boldsymbol{\tilde a}}}_{{\rm{MS}}}}({\theta _l})}
    \right|}^2}}}{{{{\left\| {{{{\boldsymbol{\hat a}}}_l}} \right\|}^2}{{\left\|
    {{{{\boldsymbol{\tilde a}}}_{{\rm{MS}}}}({\theta _l})} \right\|}^2}}}
\end{align}
The proof is completed here.

\section{} \label{appB}
For a uniform linear array, the steering vector
$\boldsymbol{a}_{\text{MS}}(\theta_i)$ can be written as
\begin{align}
\boldsymbol{a}_{\text{MS}}(\theta_{i})
\triangleq[1\phantom{0}e^{j(2\pi/\lambda)d\text{sin}(\theta_{i})}
\phantom{0}\ldots \phantom{0}
e^{j(N_{\text{MS}}-1)(2\pi/\lambda)d\text{sin}(\theta_i)}]^T
\nonumber
\end{align}
where $\lambda$ is the signal wavelength, and $d$ denotes the
distance between neighboring antenna elements. We assume each
entry of $\boldsymbol{Q}\in\mathbb{C}^{N_{\text{MS}}\times M}$ is
chosen uniformly from a unit circle scaled by a constant
$1/N_{\text{MS}}$, i.e.
$q_{m,n}=(1/N_{\text{MS}})e^{j\vartheta_{m,n}}$, where
$\vartheta_{m,n}\in [-\pi,\pi]$ follows a uniform distribution.
Let $a_{m,i}\triangleq
\boldsymbol{q}_m^T\boldsymbol{a}_{\text{MS}}(\theta_i)$ denote the
$(m,i)$th entry of $\boldsymbol{A}$, in which $\boldsymbol{q}_m$
denotes the $m$th column of $\boldsymbol{Q}$. It can be readily
verified that ${\mathbb E}[a_{m,i}]=0,\forall m,i$ and
\begin{equation}
{\mathbb E}[a_{m,i}a_{n,j}^{\ast}]=\begin{cases}0 & m\neq n \\
\frac{1}{N_{\text{MS}}^2}\boldsymbol{a}_{\text{MS}}^H(\theta_j)\boldsymbol{a}_{\text{MS}}(\theta_i)
& m=n
\end{cases}
\end{equation}
When the number of antennas at the MS is sufficiently large, the
steering vectors $\{\boldsymbol{a}_{\text{MS}}(\theta_i)\}$ become
mutually quasi-orthogonal, i.e.
$(1/N_{\text{MS}})\boldsymbol{a}_{\text{MS}}^H(\theta_j)\boldsymbol{a}_{\text{MS}}(\theta_j)\rightarrow
\delta(\theta_i-\theta_j)$, which implies that the entries of
$\boldsymbol{A}$ are uncorrelated with each other. On the other
hand, according to the central limit theorem, we know that each
entry $a_{m,i}$ approximately follows a Gaussian distribution.
Therefore entries of $\boldsymbol{A}$ can be considered as i.i.d.
Gaussian variables with zero mean and variance $1/N_{\text{MS}}$.
Thus we can reach that the k-rank of $\boldsymbol{A}$ is
equivalent to the number of columns or the number of rows,
whichever is smaller, with probability one.


\section{The Derivation of Cram\'{e}r Rao Lower Bound}\label{appC}
Consider the $M\times T\times K$ observation tensor
$\boldsymbol{\mathcal{Y}}$ in (\ref{Y-tensor})
\begin{align}
\boldsymbol{\mathcal{Y}}=\sum_{l=1}^L\alpha_l
\boldsymbol{\tilde{a}}_{\text{MS}}(\theta_{l})\circ
\boldsymbol{\tilde{a}}_{\text{BS}}(\phi_{l})\circ\boldsymbol{g}(\tau_l)
+ \boldsymbol{\mathcal{W}} \label{Y-tensor2}
\end{align}
where $\{\alpha_l,\theta_{l},\phi_{l},\tau_l\}$ are the unknown
channel parameters to be estimated. We assume that entries of
$\boldsymbol{\mathcal{W}}$ are i.i.d zero mean, circular symmetric
Gaussian random variables with variance $\sigma^2$. For ease of
exposition, let ${\boldsymbol \theta}\triangleq [\theta_1
\phantom{0} \cdots \phantom{0}\theta_L]^T$, ${\boldsymbol
\phi}\triangleq [\phi_1 \phantom{0} \cdots \phantom{0}\phi_L]^T$,
${\boldsymbol \tau}\triangleq [\tau_1 \phantom{0} \cdots
\phantom{0}\tau_L]^T$, ${\boldsymbol \alpha}\triangleq [\alpha_1
\phantom{0} \cdots \phantom{0}\alpha_L]^T$, and ${\boldsymbol{p}}
\triangleq [\boldsymbol{\theta}^T \phantom{0}\boldsymbol{\phi
}^T\phantom{0}\boldsymbol{\tau}^T\phantom{0} \boldsymbol{\alpha
}^T]$. Thus, the log-likelihood function of $\boldsymbol p$ can be
expressed as
\begin{align}
L({\boldsymbol{p}})=f({\boldsymbol {\mathcal
Y}};{\boldsymbol{A}},{\boldsymbol{B}},{\boldsymbol{C}})
\end{align}
where ${\boldsymbol A}$, ${\boldsymbol B}$ and ${\boldsymbol C}$,
defined in (\ref{factor-matrix-A}), (\ref{factor-matrix-B}) and
(\ref{factor-matrix-C}) respectively, are functions of the
parameter vector $\boldsymbol{p}$, and $f({\boldsymbol {\mathcal
Y}};{\boldsymbol{A}},{\boldsymbol{B}},{\boldsymbol{C}})$ is given
by
\begin{align*}
&f({\boldsymbol {\mathcal Y}};{\boldsymbol{A}},{\boldsymbol{B}},{\boldsymbol{C}})\\
&=  - MTK\ln (\pi {\sigma ^2}) - \frac{1}{{{\sigma ^2}}}\left\|
{{\boldsymbol{Y}}_{(1)}^T -
({\boldsymbol{C}} \odot {\boldsymbol{B}}){{\boldsymbol{A}}^T}} \right\|_F^2\\
&=  - MTK\ln (\pi {\sigma ^2}) - \frac{1}{{{\sigma ^2}}}\left\|
{{\boldsymbol{Y}}_{(2)}^T -
({\boldsymbol{C}} \odot {\boldsymbol{A}}){{\boldsymbol{B}}^T}} \right\|_F^2\\
&=  - MTK\ln (\pi {\sigma ^2}) - \frac{1}{{{\sigma ^2}}}\left\|
{{\boldsymbol{Y}}_{(3)}^T -
({\boldsymbol{B}} \odot {\boldsymbol{A}}){{\boldsymbol{C}}^T}} \right\|_F^2\\
\end{align*}
The complex Fisher information matrix (FIM) for $\boldsymbol p$ is
given by \cite{Kay93,LiuSidiropoulos01}
\begin{align}
{\boldsymbol{\Omega }}({\boldsymbol{p}}) = {\mathbb E}\left\{
{{{\left(\frac{{\partial L({\boldsymbol{p}})}}{{\partial
{\boldsymbol{p}}}}\right)}^H}\left(\frac{{\partial
L({\boldsymbol{p}})}}{{\partial {\boldsymbol{p}}}}\right)}
\right\}.
\end{align}
In the next, to calculate ${\boldsymbol{\Omega
}}({\boldsymbol{p}})$, we first compute the partial derivative of
$L({\boldsymbol p})$ with respect to $\boldsymbol p$ and then
calculate the expectation with respect to
$p(\boldsymbol{\mathcal{Y}};\boldsymbol{p})$.

\subsection{Partial Derivative of $L({\boldsymbol p})$ W.R.T $\boldsymbol{p}$}
The partial derivative of $L({\boldsymbol p})$ with respect to
$\theta_l$ can be computed as
\begin{align}
\nonumber \frac{{\partial L({\boldsymbol p})}}{{\partial {\theta
_l}}}= \text{tr}\left\{ {{{\left(\frac{{\partial L({\boldsymbol
p})}}{{\partial {\boldsymbol{A}}}}\right)}^T}\frac{{\partial
{\boldsymbol{A}}}} {{\partial {\theta _l}}}} +
{{{\left(\frac{{\partial L({\boldsymbol p})}}{{\partial
{{\boldsymbol{A}}^*}}}\right)}^T}
\frac{{\partial {{\boldsymbol{A}}^*}}}{{\partial {\theta _l}}}} \right\}\\
\end{align}
where
\begin{align}
 {{\frac{{\partial L({\boldsymbol p})}}{{\partial
{\boldsymbol{A}}}}}}=& \frac{1}{{{\sigma
^2}}}{({\boldsymbol{Y}}_{(1)}^T - ({\boldsymbol{C}} \odot
{\boldsymbol{B}}){{\boldsymbol{A}}^T})^H}({\boldsymbol{C}} \odot
{\boldsymbol{B}}) \nonumber\\
{{\frac{{\partial L({\boldsymbol p})}}{{\partial
{\boldsymbol{A}}^*}}}}=&\left({{\frac{{\partial L({\boldsymbol
p})}}{{\partial {\boldsymbol{A}}}}}}\right)^{*} \nonumber\\
\frac{{\partial {\boldsymbol{A}}}}{{\partial {\theta_l}}} =&
\left[ {\begin{array}{*{20}{c}} {\boldsymbol{0}}& \cdots
&{{\boldsymbol{\tilde a}}_l}& \cdots &{\boldsymbol{0}}
\end{array}} \right]
\end{align}
For a uniform linear array with the element spacing equal to half
of the signal wavelength, we have ${{\boldsymbol{\tilde
a}}_l}\triangleq j{
{\boldsymbol{Q}}^T{{\boldsymbol{D}}_{{a}}}{{\boldsymbol{a}}_{\rm
MS}(\theta_l)}}$, and
\begin{align}
{{\boldsymbol{D}}_{{a}}} \triangleq \pi \cos(\theta_l){\text
{diag}}(0,1, \cdots, N_{\text{MS}}-1)
\end{align}
Therefore, we have
\begin{align}
\nonumber \frac{{\partial L({\boldsymbol p})}}{{\partial {\theta
_l}}}
=& {\boldsymbol{e}}_l^T\frac{1}{{{\sigma ^2}}}{({\boldsymbol{C}} \odot {\boldsymbol{B}})^T}{({\boldsymbol{Y}}_{(1)}^T -
({\boldsymbol{C}} \odot {\boldsymbol{B}}){{\boldsymbol{A}}^T})^*}{{{\boldsymbol{\tilde a}}}_l}\\
\nonumber
&+ {\boldsymbol{e}}_l^T\frac{1}{{{\sigma ^2}}}{({\boldsymbol{C}} \odot {\boldsymbol{B}})^H}({\boldsymbol{Y}}_{(1)}^T -
({\boldsymbol{C}} \odot {\boldsymbol{B}}){{\boldsymbol{A}}^T}){\boldsymbol{\tilde a}}_l^*\\
\nonumber
= & 2{\mathop{\rm Re}\nolimits} \{ {\boldsymbol{e}}_l^T\frac{1}{{{\sigma ^2}}}{({\boldsymbol{C}} \odot
{\boldsymbol{B}})^T}{({\boldsymbol{Y}}_{(1)}^T - ({\boldsymbol{C}} \odot {\boldsymbol{B}}){{\boldsymbol{A}}^T})^*}
{{{\boldsymbol{\tilde a}}}_l}\}\\
= &2{\mathop{\rm Re}\nolimits} \{
{\boldsymbol{e}}_l^T\frac{1}{{{\sigma ^2}}}{({\boldsymbol{C}}
\odot {\boldsymbol{B}})^T}{({\boldsymbol{Y}}_{(1)}^T -
({\boldsymbol{C}} \odot
{\boldsymbol{B}}){{\boldsymbol{A}}^T})^*}{{{\boldsymbol{\tilde
A}}}{\boldsymbol{e}}_l}\}
\end{align}
where ${\rm Re}\{\cdot\}$ is an operator which takes the real part
of a complex number, ${\boldsymbol e}_l\in{\mathbb C}^{L\times 1}$
is the canonical vector whose non-zero entry is indexed as $l$,
and
\begin{align}
{\boldsymbol{\tilde A}} \triangleq \left[ {\begin{array}{*{20}{c}}
{{{{\boldsymbol{\tilde a}}}_1}}&{{{{\boldsymbol{\tilde a}}}_2}}&
\cdots &{{{{\boldsymbol{\tilde a}}}_L}}
\end{array}} \right]
\end{align}
Similarly, we can obtain the partial derivatives with respect to
other parameters as follows
\begin{align}
\nonumber
&\frac{{\partial L(\boldsymbol p)}}{{\partial {\phi _l}}} = 2{\mathop{\rm Re}\nolimits} \{ {\boldsymbol{e}}_l^T\frac{1}
{{{\sigma ^2}}}{({\boldsymbol{C}} \odot {\boldsymbol{A}})^T}{({\boldsymbol{Y}}_{(2)}^T -
({\boldsymbol{C}} \odot {\boldsymbol{A}}){{\boldsymbol{B}}^T})^*}{{\boldsymbol{\tilde B}}{{\boldsymbol e}_l}}\}\\
\nonumber &\frac{{\partial L({\boldsymbol{p}})}}{{\partial {\tau
_l}}} = 2{\mathop{\rm Re}\nolimits} \{
{\boldsymbol{e}}_l^T\frac{1}{{{\sigma ^2}}}{({\boldsymbol{B}}
\odot {\boldsymbol{A}})^T}{({\boldsymbol{Y}}_{(3)}^T -
({\boldsymbol{B}} \odot {\boldsymbol{A}}){{\boldsymbol{C}}^T})^*}{{\boldsymbol{\tilde C}}{\boldsymbol e}_l}\}\\
\nonumber &\frac{{\partial L({\boldsymbol{p }})}}{{\partial
{\alpha _l}}} = {\boldsymbol{e}}_l^T\frac{1}{{{\sigma
^2}}}{({\boldsymbol{B}} \odot
{\boldsymbol{A}})^T}{({\boldsymbol{Y}}_{(3)}^T - ({\boldsymbol{B}}
\odot
{\boldsymbol{A}}){{\boldsymbol{C}}^T})^*}{{\boldsymbol{G}}{\boldsymbol
e}_l}
\end{align}
where
\begin{align}
{\boldsymbol{\tilde B}} &\triangleq \left[
{\begin{array}{*{20}{c}} {{{{\boldsymbol{\tilde
b}}}_1}}&{{{{\boldsymbol{\tilde b}}}_2}}& \cdots
&{{{{\boldsymbol{\tilde b}}}_L}}
\end{array}} \right]\\
{\boldsymbol{\tilde C}} &\triangleq \left[
{\begin{array}{*{20}{c}} {{{{\boldsymbol{\tilde
c}}}_1}}&{{{{\boldsymbol{\tilde c}}}_2}}& \cdots
&{{{{\boldsymbol{\tilde c}}}_L}}
\end{array}} \right]\\
{\boldsymbol{ G}} &\triangleq \left[ {\begin{array}{*{20}{c}}
{{{{\boldsymbol{ g}}}_1}}&{{{{\boldsymbol{ g}}}_2}}& \cdots
&{{{{\boldsymbol{ g}}}_L}}
\end{array}} \right]
\end{align}
in which ${{\boldsymbol{\tilde b}}_l}\triangleq j{
{\boldsymbol{P}}^T {\boldsymbol D}_{b}{{\boldsymbol a}_{\rm
BS}}(\phi_l)}$, ${\boldsymbol{\tilde c}}_l \triangleq
j{{\boldsymbol{D}}_{{c}}}{{\boldsymbol{c}}_l}$, and
\begin{align}
{\boldsymbol D}_{b} &\triangleq \pi \cos(\phi_l){\text {diag}}(
0,1, \cdots, N_{\text{BS}}-1)\\
{{\boldsymbol{D}}_{c}} &\triangleq -2\pi {\text
{diag}}(0,f_s/\bar{K}, \cdots, (K-1)f_s/\bar{K})
\end{align}

\subsection{Calculation of Fisher Information Matrix}
We first calculate the entries in the principal minors of
$\boldsymbol{\Omega}(\boldsymbol{p})$. For instance, the
$(l_1,l_2)$th entry of
\begin{align*}
{\mathbb E}\left\{ {{{\left(\frac{{\partial
L({\boldsymbol{p}})}}{{\partial
{\boldsymbol{\theta}}}}\right)}^H}\left(\frac{{\partial
L({\boldsymbol{p}})}}{{\partial {\boldsymbol{\theta}}}}\right)}
\right\}.
\end{align*}
is given by
\begin{align}
\nonumber
&{\mathbb E}\left\{ {{{\left(\frac{{\partial L({\boldsymbol{p}})}}{{\partial {\theta _{{l_1}}}}}\right)}^*}
\left(\frac{{\partial L({\boldsymbol{p}})}}{{\partial {\theta _{{l_2}}}}}\right)} \right\}\\
\nonumber
&= 4{\mathbb E}\left[ {{\mathop{\rm Re}\nolimits} \{ {\boldsymbol{e}}_{{l_1}}^T{{{\boldsymbol{N}}}^{ a}}
{{\boldsymbol{e}}_{{l_1}}}\} {\mathop{\rm Re}\nolimits} \{ {\boldsymbol{e}}_{{l_2}}^T{{{\boldsymbol{ N}}}^{ a}}
{{\boldsymbol{e}}_{{l_2}}}\} } \right]\\
\nonumber &= {\mathbb E}\left[ {\left({{{\boldsymbol{
N}}}^{{a}}}({l_1},{l_1}) + {{{\boldsymbol{
N}}}^{{a}}}{{({l_1},{l_1})}^*}\right)\left({{{\boldsymbol{
N}}}^{{a}}}({l_2},{l_2}) + {{{\boldsymbol{
N}}}^{{a}}}{{({l_2},{l_2})}^*}\right)} \right]
\end{align}
where ${{{\boldsymbol{ N}}}^{{a}}}({l_1},{l_1})$ stands for the
$(l_1,l_1)$th entry of ${{{\boldsymbol{ N}}}^{{a}}}\in {\mathbb
C}^{L\times L}$ and
\begin{align}
\nonumber
{{{\boldsymbol{ N}}}^{a}}&\triangleq\frac{1}{{{\sigma ^2}}}{({\boldsymbol{C}} \odot {\boldsymbol{B}})^T}
{({\boldsymbol{Y}}_{(1)}^T - ({\boldsymbol{C}} \odot \boldsymbol{B}){{\boldsymbol{A}}^T})^*}{\boldsymbol{\tilde A}}\\
\nonumber &=\frac{1}{{{\sigma ^2}}}{({\boldsymbol{C}} \odot
{\boldsymbol{B}})^T}{({{\boldsymbol W}^T_{(1)}})^*}{\boldsymbol{
\tilde A}}.
\end{align}
Letting ${{{\boldsymbol{
n}}}^{a}}\triangleq\text{vec}({{{\boldsymbol{ N}}}^{a}})$, we have
\begin{align}
{{{\boldsymbol{ n}}}^{a}}=\left({\boldsymbol{ \tilde
A}}^T\otimes{({\boldsymbol{C}} \odot
{\boldsymbol{B}})^T}\right){\text {vec}} ({{\boldsymbol
W}^H_{(1)}}).
\end{align}
where ${{\boldsymbol W}_{(1)}}$ is the mode-1 unfolding of
$\boldsymbol{\mathcal{W}}$, thus ${\text{vec}}({{\boldsymbol
W}^H_{(1)}})$ is a zero mean circularly symmetric complex Gaussian
vector whose covariance matrix is given by
$\sigma^2\boldsymbol{I}$. Since ${{{\boldsymbol{ n}}}^{a}}$ is the
linear transformation of ${\text{vec}}({{\boldsymbol
W}^H_{(1)}})$, ${{{\boldsymbol{ n}}}^{a}}$ also follows a
circularly symmetric complex Gaussian distribution. Its covariance
matrix ${\boldsymbol C}_{{{\boldsymbol{ n}}}^{ a}}\in{\mathbb
C}^{L^2\times L^2}$ and second-order moments ${\boldsymbol
M}_{{{\boldsymbol{ n}}}^{ a}}\in{\mathbb C}^{L^2\times L^2}$ are
respectively given by
\begin{align}
\nonumber
\boldsymbol{C}_{\boldsymbol{n}^{a}}&={\mathbb E}\left[(\boldsymbol{n}^{a})(\boldsymbol{n}^{a})^H\right]\\
\nonumber
&=\frac{1}{\sigma^2}\left(\boldsymbol{\tilde{A}}^T\otimes{({\boldsymbol{C}}
\odot {\boldsymbol{B}})^T}\right)
\left(\boldsymbol{\tilde{A}}^*\otimes{(\boldsymbol{C} \odot \boldsymbol{B})^*}\right)\\
&=\frac{1}{{{\sigma
^2}}}\left(\boldsymbol{\tilde{A}}^T\boldsymbol{\tilde{
A}}^*\right)\otimes\left({(\boldsymbol{C} \odot
\boldsymbol{B})^T}{(\boldsymbol{C}\odot\boldsymbol{B})^*}\right)
\end{align}
and
\begin{align}
{\boldsymbol M}_{{{\boldsymbol{ n}}}^{ a}}={\mathbb E}\left[
({{{\boldsymbol{ n}}}^{ a}})({{{\boldsymbol{ n}}}^{ a}})^T
\right]= \boldsymbol{0}
\end{align}
Therefore, we have
\begin{align}
&{\mathbb E}\left\{ {{{\left(\frac{{\partial
L(\boldsymbol{p})}}{{\partial {\theta
_{{l_1}}}}}\right)}^*}\left(\frac{{\partial
L(\boldsymbol{p})}}{{\partial {\theta _{{l_2}}}}}\right)}
\right\}=2{\rm Re}\{{\boldsymbol C}_{{{\boldsymbol{n}}}^{
a}}(m,n)\}
\end{align}
where $m\triangleq L(l_1-1)+l_1$ and $n\triangleq L(l_2-1)+l_2$.
Similarly, we can arrive at
\begin{align*}
{\mathbb E}\left\{ {{{\left(\frac{{\partial
L({\boldsymbol{p}})}}{{\partial {\phi _{{l_1}}}}}\right)}^*}
\left(\frac{{\partial L({\boldsymbol{p}})}}{{\partial {\phi
_{{l_2}}}}}\right)} \right\}&=
2{\mathop{\rm Re}\nolimits} \{ {{\boldsymbol{C}}_{{{{\boldsymbol{ n}}}^{ b}}}}(m,n)\} \\
{\mathbb E}\left\{ {{{\left(\frac{{\partial
L({\boldsymbol{p}})}}{{\partial {\tau _{{l_1}}}}}\right)}^*}
\left(\frac{{\partial L(\boldsymbol{p})}}{{\partial {\tau
_{{l_2}}}}}\right)} \right\}&=
2{\mathop{\rm Re}\nolimits} \{ {\boldsymbol{C}}_{{{{\boldsymbol{ n}}}^{ c}}}(m,n)\}\\
{\mathbb E}\left\{ {{{\left(\frac{{\partial
L({\boldsymbol{p}})}}{{\partial {\alpha
_{{l_1}}}}}\right)}^*}\left(\frac{{\partial
L(\boldsymbol{p})}}{{\partial{\alpha _{{l_2}}}}}\right)}
\right\}&= {\boldsymbol{C}}_{{{{\boldsymbol{\tilde
n}}}^{c}}}(m,n)^*,
\end{align*}
in which
\begin{align}
{{\boldsymbol{C}}_{{{{\boldsymbol{ n}}}^{
b}}}}\triangleq\frac{1}{{{\sigma ^2}}}
\left(\boldsymbol{\tilde{B}}^T\boldsymbol{\tilde{B}}^{*}\right)\otimes
\left({(\boldsymbol{C}\odot\boldsymbol{A})^T}{(\boldsymbol{C}\odot\boldsymbol{A})^{*}}\right)\\
{{\boldsymbol{C}}_{{{{\boldsymbol{ n}}}^{
c}}}}\triangleq\frac{1}{{{\sigma^2}}}
\left(\boldsymbol{\tilde{C}}^T\boldsymbol{\tilde{C}}^{*}\right)
\otimes\left({(\boldsymbol{B}\odot\boldsymbol{A})^T}{(\boldsymbol{B}\odot\boldsymbol{A})^*}\right)\\
{{\boldsymbol{C}}_{{{{\boldsymbol{ \tilde n}}}^{
c}}}}\triangleq\frac{1}{{{\sigma
^2}}}\left(\boldsymbol{G}^T\boldsymbol{G}^*\right)\otimes\left({(\boldsymbol{B}
\odot\boldsymbol{A})^T}{(\boldsymbol{B}\odot\boldsymbol{A})^{*}}\right)
\end{align}
For the elements in the off-principal minors of
$\boldsymbol{\Omega}(\boldsymbol{p})$, such as the $(l_1,l_2)$th
entry of
\begin{align*}
{\mathbb E}\left\{ {{{\left(\frac{{\partial
L({\boldsymbol{p}})}}{{\partial
{\boldsymbol{\theta}}}}\right)}^H}\left(\frac{{\partial
L({\boldsymbol{p}})}}{{\partial {\boldsymbol{\phi}}}}\right)}
\right\}.
\end{align*}
is given by
\begin{align*}
&{\mathbb E}\left\{ {{{\left(\frac{{\partial
L({\boldsymbol{p}})}}{{\partial {\theta _{{l_1}}}}}\right)}^*}
\left(\frac{{\partial L({\boldsymbol{p}})}}{{\partial {\phi _{{l_2}}}}}\right)} \right\}\\
&=4{\mathbb E}\left[ {{\mathop{\rm Re}\nolimits} \left\{
{\bf{e}}_{{l_1}}^T{{{\boldsymbol{ N}}}^{
a}}{{\bf{e}}_{{l_1}}}\right\}
{\mathop{\rm Re}\nolimits} \{ {\bf{e}}_{{l_2}}^T{{{\boldsymbol{ N}}}^{ b}}{{\boldsymbol{e}}_{{l_2}}}\} } \right]\\
&= {\mathbb E}\left[ {({{{\boldsymbol{ N}}}^{{a}}}({l_1},{l_1}) +
{{{\boldsymbol{ N}}}^{{a}}}{{({l_1},{l_1})}^*})({{{\boldsymbol{
N}}}^{{b}}}({l_2},{l_2}) +
{{{\boldsymbol{ N}}}^{{b}}}{{({l_2},{l_2})}^*})} \right]\\
&= 2{\mathop{\rm Re}\nolimits}\left \{
{\boldsymbol{C}}_{{{{{\boldsymbol{ n}}}^{ a}}},{{{{\boldsymbol{
n}}}^{ b}}}}(m,n)\right\}
\end{align*}
where
\begin{align}
\nonumber {\boldsymbol{C}}_{{{{{\boldsymbol{ n}}}^{
a}}},{{{{\boldsymbol{ n}}}^{ b}}}} &\triangleq
{\mathbb E}\left[(\boldsymbol{n}^{a})(\boldsymbol{n}^{b})^H\right]\\
&= \frac{1}{{{\sigma ^4}}}{({\boldsymbol{\tilde A}} \otimes
({\boldsymbol{C}} \odot
{\boldsymbol{B}}))^T}{\boldsymbol{C}}_{{{\boldsymbol
w}_1},{{\boldsymbol w}_2}} ({{{\boldsymbol{\tilde B}}}^*} \otimes
{({\boldsymbol{C}} \odot {\boldsymbol{A}})^*})
\end{align}
in which
\begin{align}
{\boldsymbol{C}}_{{{\boldsymbol w}_1},{{\boldsymbol
w}_2}}\triangleq{\mathbb E}\{ {\text
{vec}}({\boldsymbol{W}}_{(1)}^H){\text
{vec}}{({\boldsymbol{W}}_{(2)}^T)^T}\}
\end{align}
Similarly, we can obtain
\begin{align*}
{\mathbb E}\left\{ {{{\left(\frac{{\partial
f(\boldsymbol{p})}}{{\partial {\theta _{{l_1}}}}}\right)}^*}
\left(\frac{{\partial f(\boldsymbol{p})}}{{\partial {\tau
_{{l_2}}}}}\right)} \right\}&=
2{\rm {Re}}\{ \boldsymbol{C}_{\boldsymbol{n}^a,\boldsymbol{n}^c}(m,n)\}\\
{\mathbb E}\left\{ {{{\left(\frac{{\partial
f(\boldsymbol{p})}}{{\partial {\theta _{{l_1}}}}}\right)}^*}
\left(\frac{{\partial f(\boldsymbol{p})}}{{\partial {\alpha
_{{l_2}}}}}\right)} \right\}&=
\boldsymbol{C}_{{\boldsymbol n}^a,{\boldsymbol {\tilde n}}^c}(m,n)^*\\
{\mathbb E}\left\{ {{{\left(\frac{{\partial
f(\boldsymbol{p})}}{{\partial{\phi _{{l_1}}}}}\right)}^*}
\left(\frac{{\partial f(\boldsymbol{p})}}{{\partial {\tau
_{{l_2}}}}}\right)} \right\}&=
2{\rm {Re}}\{ \boldsymbol{C}_{{\boldsymbol n}^b,{\boldsymbol n}^c}(m,n)\}\\
{\mathbb E}\left\{ {{{\left(\frac{{\partial
f(\boldsymbol{p})}}{{\partial{\phi _{{l_1}}}}}\right)}^*}
\left(\frac{{\partial f(\boldsymbol{p})}}{{\partial {\alpha
_{{l_2}}}}}\right)} \right\}&=
{{\boldsymbol{C}}_{{\boldsymbol n}^b,{\boldsymbol {\tilde n}}^c}}(m,n)^*\\
{\mathbb E}\left\{ {{{\left(\frac{{\partial
f(\boldsymbol{p})}}{{\partial {\tau
_{{l_1}}}}}\right)}^*}\left(\frac{{\partial
f(\boldsymbol{p})}}{{\partial{\alpha _{{l_2}}}}}\right)}
\right\}&= \boldsymbol{C}_{{\boldsymbol n}^c,{\boldsymbol {\tilde
n}}^c}(m,n)^*
\end{align*}
where
\begin{align*}
\nonumber &{\boldsymbol{C}}_{{{{{\boldsymbol{ n}}}^{
a}}},{{{{\boldsymbol{ n}}}^{ c}}}} \triangleq\frac{1}{{{\sigma
^4}}}{({\boldsymbol{\tilde A}} \otimes ({\boldsymbol{C}} \odot
{\boldsymbol{B}}))^T}{\boldsymbol{C}}_{{{\boldsymbol
w}_1},{{\boldsymbol w}_3}}
({{{\boldsymbol{\tilde C}}}^*} \otimes {({\boldsymbol{B}} \odot {\boldsymbol{A}})^*})\\
&{\boldsymbol{C}}_{{{{{\boldsymbol{ n}}}^{
a}}},{{{{\boldsymbol{\tilde  n}}}^{ c}}}}
\triangleq\frac{1}{{{\sigma ^4}}}{({\boldsymbol{\tilde A}} \otimes
({\boldsymbol{C}} \odot
{\boldsymbol{B}}))^T}{\boldsymbol{C}}_{{{\boldsymbol
w}_1},{{\boldsymbol w}_3}}
({{{\boldsymbol{ G}}}^*} \otimes {({\boldsymbol{B}} \odot {\boldsymbol{A}})^*})\\
&{\boldsymbol{C}}_{{{{{\boldsymbol{ n}}}^{ b}}},{{{{\boldsymbol{
n}}}^{ c}}}} \triangleq\frac{1}{{{\sigma
^4}}}{({\boldsymbol{\tilde B}} \otimes ({\boldsymbol{C}} \odot
{\boldsymbol{A}}))^T}{\boldsymbol{C}}_{{{\boldsymbol
w}_2},{{\boldsymbol w}_3}}
({{{\boldsymbol{\tilde C}}}^*} \otimes {({\boldsymbol{B}} \odot {\boldsymbol{A}})^*})\\
&{\boldsymbol{C}}_{{{{{\boldsymbol{ n}}}^{
b}}},{{{{\boldsymbol{\tilde n}}}^{ c}}}}
\triangleq\frac{1}{{{\sigma ^4}}}{({\boldsymbol{\tilde B}} \otimes
({\boldsymbol{C}} \odot
{\boldsymbol{A}}))^T}{\boldsymbol{C}}_{{{\boldsymbol
w}_2},{{\boldsymbol w}_3}}
({{{\boldsymbol{ G}}}^*} \otimes {({\boldsymbol{B}} \odot {\boldsymbol{A}})^*})\\
&{\boldsymbol{C}}_{{{{{\boldsymbol{ n}}}^{ c}}},{{{{\boldsymbol{
\tilde n}}}^{ c}}}} \triangleq\frac{1}{{{\sigma
^2}}}{({\boldsymbol{\tilde C}} \otimes ({\boldsymbol{B}} \odot
{\boldsymbol{A}}))^T} ({{{\boldsymbol{ G}}}^*} \otimes
{({\boldsymbol{B}} \odot {\boldsymbol{A}})^*})
\end{align*}
in which
\begin{align*}
\boldsymbol{C}_{\boldsymbol{w}_1,\boldsymbol{w}_3}&\triangleq{\mathbb
E}\{\text{vec}
({\boldsymbol{W}}_{(1)}^H){\text {vec}}{({\boldsymbol{W}}_{(3)}^T)^T}\}\\
\boldsymbol{C}_{{\boldsymbol w}_2,{\boldsymbol
w}_3}&\triangleq{\mathbb E}\{ {\text
{vec}}({\boldsymbol{W}}_{(2)}^H){\text
{vec}}{({\boldsymbol{W}}_{(3)}^T)^T}\}.
\end{align*}

The computation of $\boldsymbol{C}_{{\boldsymbol w}_1,{\boldsymbol
w}_2}$ is elaborated as follows. Note that the $(m,t,k)$th entry
in $\boldsymbol {\mathcal W}\in {\mathbb C}^{M\times T\times K}$
corresponds to the $(m,t+(k-1)T)$th entry of ${\boldsymbol
W}_{(1)}$ and also corresponds to the $(t,m+(k-1)M)$th entry of
${\boldsymbol W}_{(2)}$. Furthermore, the $(m,t+(k-1)T)$th entry
of ${\boldsymbol W}_{(1)}$ corresponds to the
$(t+(k-1)T+(m-1)TK)$th entry of ${{{\text
{vec}}}({{{\boldsymbol{W}}}^H_{(1)}})}$ and the $(t,m+(k-1)M)$th
entry of ${\boldsymbol W}_{(2)}$ corresponds to the
$(m+(k-1)M+(t-1)MK)$th entry of ${{{\text
{vec}}}({{{\boldsymbol{W}}}^T_{(2)}})}$. Since entries in
$\boldsymbol {\mathcal W}$ are i.i.d. random variables, i.e.,
\begin{align*}
{\mathbb E}\{ {w_{{m_1},{t_1},{k_1}}}w_{{m_2},{t_2},{k_2}}^*\}  = \left\{ {\begin{aligned}
&{{\sigma ^2};{m_1} = {m_2},{t_1} = {t_2},{k_1} = {k_2}}\\
&{0;\text{otherwise}}
\end{aligned}} \right.
\end{align*}
where ${w_{{m},{t},{k}}}$ represents the $(m,t,k)$th entry of
$\boldsymbol {\mathcal W}$. Therefore, in
$\boldsymbol{C}_{{\boldsymbol w}_1,{\boldsymbol w}_2}\in {\mathbb
C}^{TKM \times MKT}$, the number of nonzero entries is $MTK$ and
the corresponding indexes,
$\{(n_1,n_2)|\boldsymbol{C}_{{\boldsymbol w}_1,{\boldsymbol
w}_2}(n_1,n_2)\neq0\}$, is equal to
\begin{align*}
\{(n_1,n_2)|&n_1 = t + (k - 1)T + (m - 1)TK,\\
&n_2 = m + (k - 1)M + (t - 1)MK, \forall m,t,k \}
\end{align*}

Similarly, the index of the nonzero elements in
$\boldsymbol{C}_{{\boldsymbol w}_1,{\boldsymbol w}_3}\in {\mathbb
C}^{TKM \times MTK}$ and $\boldsymbol{C}_{{\boldsymbol
w}_2,{\boldsymbol w}_3}\in {\mathbb C}^{MKT \times MTK}$ are
respectively belongs to
\begin{align*}
 \{(n_1,n_2)|&n_1 = t+(k-1)T+(m-1)TK,\\
&n_2 = m+(t-1)M+(k-1)MT, \forall m,t,k\}
\end{align*}
and
\begin{align*}
 \{(n_1,n_2)|&n_1 = m+(k-1)M+(t-1)MK,\\
&n_2 = m+(t-1)M+(k-1)MT, \forall m,t,k\}
\end{align*}

\subsection{Cram\'{e}r Rao bound}
After obtaining the fisher information matrix, the CRB for the
parameters $\boldsymbol p$ can be calculated as \cite{Kay93}
\begin{align}
{\rm {CRB}}(\boldsymbol{p})=
\boldsymbol{\Omega}^{-1}(\boldsymbol{p})
\end{align}

\bibliographystyle{IEEEtran}


\end{document}